\documentclass[prc,twocolumn,secnumroman,tightenlines,amssymb,aps,nobibnotes,
  superscriptaddress,showpacs,balancelastpage,floatfix,nofootinbib]{revtex4}
\usepackage{graphicx,colordvi}
\usepackage{multirow}  
\usepackage{color}	
\usepackage[utf8]{inputenc}
\usepackage{amsmath}	
\newcommand{\beq}[0]{\begin{equation}}
\newcommand{\eeq}[0]{\end{equation}}
\newcommand{\bea}[0]{\begin{eqnarray}}
\newcommand{\eea}[0]{\end{eqnarray}}

\begin{document}

\title{On the stability of super-heavy nuclei}
\author{K. Pomorski}\email{pomorski@kft.umcs.lublin.pl}
\affiliation{Uniwersytet Marii Curie Sk\l odowskiej, Katedra Fizyki
Teoretycznej, 20031 Lublin, Poland}
\author{B. Nerlo-Pomorska}
\affiliation{Uniwersytet Marii Curie Sk\l odowskiej, Katedra Fizyki
Teoretycznej, 20031 Lublin, Poland}
\author{J. Bartel}
\affiliation{IPHC, Universit\'e de Strasbourg-CNRS, 67037 Strasbourg, France}
\author{C. Schmitt}
\affiliation{IPHC, Universit\'e de Strasbourg-CNRS, 67037 Strasbourg, France}

\date{\today}
\begin{abstract}
\noindent

The potential-energy surfaces of an extended set of heavy and super-heavy
even-even nuclei with $92 \le Z \le 126$ and isospins $40 \le N \!-\!Z \le 74$
are evaluated within the recently developed Fourier shape parametrization.
Ground-state and decay properties are studied for 324 different even-even 
isotopes in a four-dimensional deformation space, defined by non-axiality,
quadrupole, octupole, and hexadecapole degrees of freedom. Nuclear deformation 
energies are evaluated in the framework of the macroscopic-microscopic approach,
with the Lublin-Strasbourg-Drop model and a Yukawa-folded mean-field potential.
The evolution of the ground-state equilibrium shape (and possible isomeric,
metastable states) is studied as a function of $Z$ and $N$. Alpha-decay
$Q$-values and half-lives, as well as fission-barrier heights, are deduced. In
order to understand the transition from asymmetric to symmetric fission along
the Fm isotopic chain, the properties of all identified fission paths are
investigated. Good agreement is found with experimental data wherever 
available. New interesting features about the population of different fission
modes for nuclei beyond Fm are predicted.
\end{abstract}
\maketitle

\section{Introduction}

The properties of nuclei at the edge of the nuclear chart and consequently at
the limit of stability, namely in the region of very-heavy (VHE) and super-heavy
(SHE) elements, are of paramount interest, since they constitute a stringent
test of any nuclear model. While many theories are nowadays indeed able to
achieve a fair description of nuclear masses for nuclei on and close to the
$\beta$-stability line, these can substantially deviate when moving to the SHE
region. Apart from the fundamental interest to achieve a better understanding of
the involved physics, the predictive power of these theoretical approaches is of
capital importance to guide the challenging experimental quest for the 
so-called SHE island of stability.
                                                                     \\[ -2.0ex]

Models used in this field can be essentially classified into two categories, 
the self-consistent microscopic approach rooted, on a more or less fundamental
level (effective interactions, meson fields, quark degrees of freedom, $\cdots$)
in the underlying nuclear force, and the macroscopic-microscopic model that
describes the nucleus as a charged liquid drop with quantal (shell and pairing)
corrections. Even though the self-consistent microscopic theory has been
developed substantially, its achievements depend strongly on the specific
nuclear interaction used (see e.g. Ref.\ \cite{bender2000}). The large
computing-resources required are, in addition, still a limiting factor for
systematic investigations. The macroscopic-microscopic approach, on the other
hand, has proven to constitute a reliable method for addressing a very wide
variety of questions in the field, and this with impressively good accuracy
\cite{yyy}.
                                                                     \\[ -2.0ex]

The present work is a continuation of our previous investigation \cite{SNP17} in
which we have introduced a new, powerful and rapidly converging description of
nuclear deformations based on a Fourier decomposition of the nuclear shape.
Combined with a well-established macroscopic-microscopic model, nuclear
deformation-energy landscapes have been investigated for preactinides and
actinides ($78 \!\le\! Z \!\le\! 94$) and shown to be in good agreement with the
available experimental data. The present study proposes to extend the
application of the model into the less-known SHE region.
                                                                     \\[ -2.0ex]

The innovative Fourier shape parametrization and the main features of the 
macroscopic-microscopic potential-energy calculation are discussed in section 
II. The results of our calculations are presented in section III, discussing 
first the equilibrium deformations and their dependence on $Z$ and $N$.
Combining the thus obtained deformation-energy landscapes with a simple 
Wentzel-Kramers-Brillouin (WKB) model \cite{ZWP13}, $\alpha$-decay properties, 
including $Q$-value and half-life, are obtained. Fission barrier heights are 
also derived from the potential-energy landscapes. Finally, the topography of 
the 4D deformation space is investigated in detail to search for most probable 
paths to fission, and their evolution with neutron and proton numbers across 
the Fm region. All along, comparison with experiment is made wherever 
measurements are available. A summary and conclusions are drawn in Sec. IV.


\section{Theoretical framework}\label{model}


\subsection{The Fourier shape parametrization}

The description of the huge variety of shapes encountered all across the 
nuclear chart, from oblate deformations found in the transition region and 
corresponding to the progressive filling of the $pf$ shell, to prolate shapes 
as realized in numerous rare-earth and actinide isotopes, requires a rich and 
flexible nuclear shape parametrization. The requirement is even more demanding 
for describing fissioning shapes, which are typically very elongated and
necked-in. To model the physical reality as faithfully as possible (as far as
that could be identified), it is desirable that the parametrization involves a 
large amount of deformation parameters, in order to take into account all of
the degrees of freedom involved. For a numerical treatment, on the other hand,
a very large number of deformation coordinates is prohibitive. 
The challenge therefore is to isolate the essential degrees of freedom and to
describe these with a few physically relevant deformation parameters. Several
shape parametrization have been proposed, and are currently used for
investigating the properties and decay of nuclei. The series expansion of the
nuclear radius in spherical harmonics proposed by Lord Rayleigh \cite{LR79} 
already in the 19th century, turned out to be able to describe a very rich
variety of shapes, and is, till nowadays, one of the most widely used 
prescription in nuclear structure calculations. The achievement of this 
expansion strongly relies on the large amount of degrees of freedom taken into 
account. As soon, however, as elongated configurations are to be modeled, and 
most crucially for fissioning shapes, a nearly prohibitive number of 
deformation parameters is required (as many as seven parameters were needed to 
describe the height of the fission barrier of $^{232}$Th when imposing a 
left-right symmetric mass split \cite{JKS13, dobrowolski:2007}). Alternative
parametrizations have been proposed since the days of Lord Rayleigh. Among the 
most powerful and popular ones we cite the quadratic surfaces of revolution
(QSR) \cite{Ni69}, the Cassini ovals \cite{Pa71,PR08}, the Funny-Hills (FH)
shapes \cite{FH} and its modified version \cite{MFH}, as well as the expansion
of the nuclear surface in a series of Legendre polynomials \cite{TKS80}. All 
these parametrizations are able to describe nuclear potential-energy landscapes
rather well, and, in most cases, with a limited number of deformation degrees 
of freedom. 
For all of them, however, except those of Refs. \cite{LR79,TKS80}, they have 
the inconvenience that they do not allow to control their convergence. In what 
follows, we are going to use an innovative parametrization, initially 
introduced in Ref.\ \cite{PPB15}, based on a Fourier expansion of the nuclear 
surface. This prescription has been shown \cite{SNP17} to be rapidly converging 
and to describe nuclear ground-state configurations, as well as very elongated
and necked-in shapes, as they are encountered in the fission process close to
scission with few deformation parameters only. 
                                   \\[-4.0ex]

Within this Fourier parametrization \cite{PPB15,SNP17}, we write, for
axially symmetric shapes (a generalization to triaxial shapes will be given
below), the distance $\rho^{}_s(z)$ from the symmetry axis of a surface point 
at coordinate $z$ in cylindrical coordinates as
\begin{widetext}
\beq
 \frac{\rho_s^2(z)}{R_{0}^2} =\! \sum\limits_{n=1}^\infty \left[
 a^{}_{2n} \cos\left(\frac{(2n-1) \pi}{2} \, \frac{z-z_{sh}}{z_0}\right) 
 + a^{}_{2n+1} \sin\left(\frac{2 n \pi}{2} \, \frac{z-z_{sh}}{z_0}\right)
         \right]
\label{Eq-02}\eeq
\end{widetext}
where $R^{}_{0}$ is the radius of the corresponding spherical shape with
the same volume. The extension of the shape along the symmetry axis is $2 z_0 =
2 c R_0$ with left and right ends located at $z_{\rm min} = z_{\rm sh} - z_0$
and $z_{\rm max} = z_{\rm sh} + z_0$, where $\rho_s^2(z)$ vanishes, a condition
which is automatically satisfied by Eq.\ (\ref{Eq-02}). Here $c = z_0/R_0$ is
the Funny-Hills \cite{FH} elongation parameter ($c<1$ for oblate, $c>1$ for
prolate shapes) which is related to the even Fourier coefficients by
the volume-conservation relation:
\beq
 \frac{\pi}{3c} = \sum\limits_{n=1}^\infty (-1)^{n-1} \,\frac{a_{2n}}{2n-1}\,\,.
\label{Eq-03}\eeq
The shift coordinate $z_{\rm sh}$ in (\ref{Eq-02}) is chosen such that the
center of mass of the nuclear shape is located at the origin of the coordinate
system. The parameters $a_2,~a_3,~a_4$ describe, respectively, quadrupole,
octupole and hexadecapole type deformations, which in the context of fission,
are related to elongation, left-right asymmetry, and neck thickness,
respectively.
                                                                     \\[ -2.0ex]

Notice that our Fourier shape parametrization is, in a way, quite similar to 
the one due to Lord Rayleigh, in the sens that it is an expansion in a complete 
basis (trigonometric functions here, spherical harmonics there), but with the 
essential difference that, instead of the nuclear radius $R(\theta,\varphi)$, 
it is now the square of the distance $\rho_s^2(z)$ of a surface point from the 
symmetry axis that is expanded in the basis functions. Such a description 
seems, indeed, to be better adapted to that kind of physical problem, as the 
success of the Funny-Hills \cite{FH,MFH}
or the Trentalange-Koonin-Sierk shapes \cite{TKS80} indicate.
                                                                     \\[ -2.0ex]

\begin{figure}[!hbt]
\hspace{-1.cm}
\includegraphics[width=7.6cm]{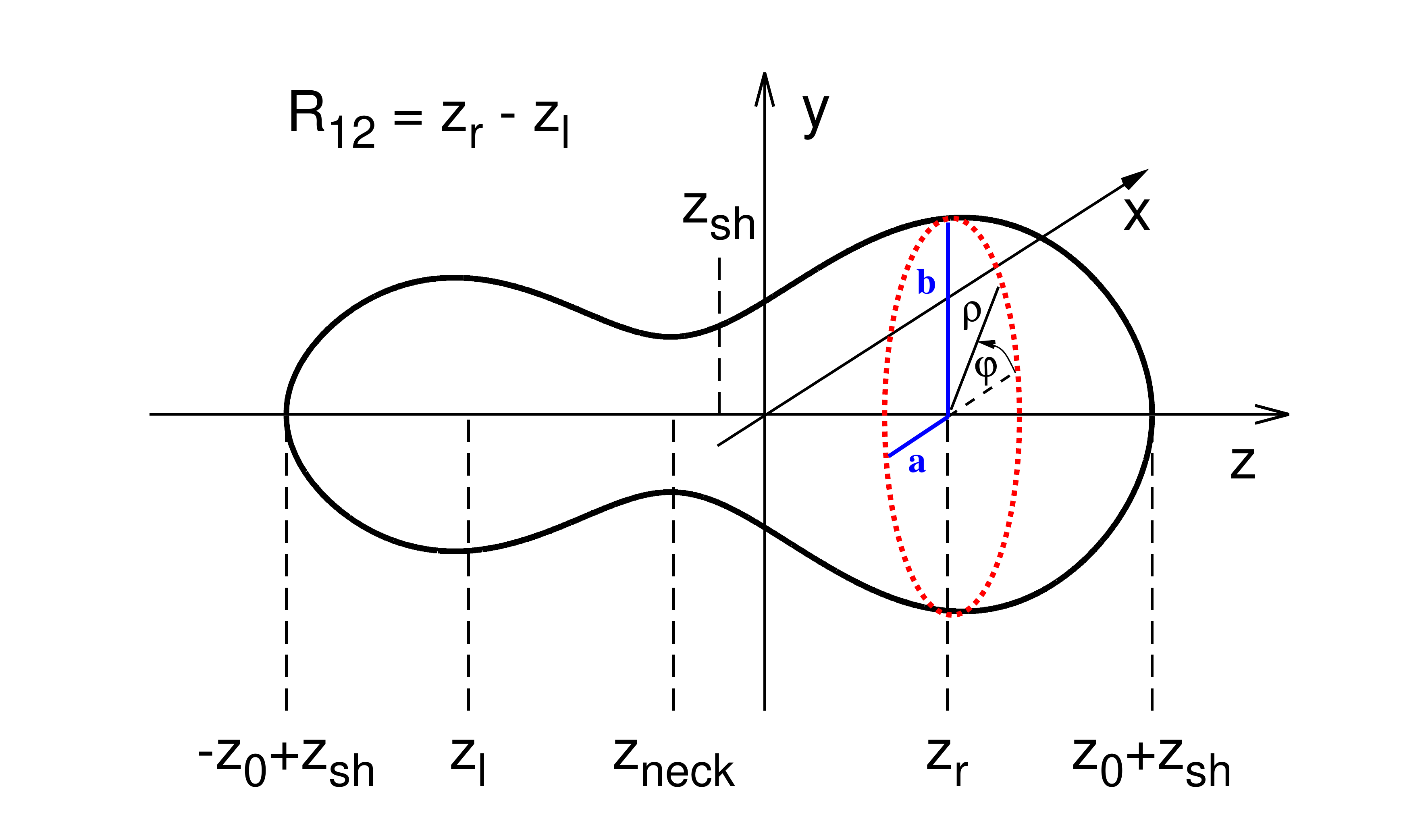}
\caption{(Color online) Schematic visualization, in cylindrical coordinates, 
of the parameters entering the definition of the profile function defined by 
Eq.\ (\ref{Eq-05}). The quantities $z_{\rm l}$ and $z_{\rm r}$ localize the 
mass centers of left and right nascent fragment entering the definition of $R_{12} = z_{\rm r} - z_{\rm l}$.} 
\label{fig01}\end{figure}

As an example, a nuclear shape typically realized in the course of the fission
process is displayed in Fig.\ \ref{fig01}. The various quantities discussed in
the text are indicated in the figure.
                                                                     \\[ -2.0ex]

To describe non-axial shapes, the cross section perpendicular to the symmetry
axis is assumed to be of ellipsoidal form, and defined by a non-axiality
parameter 
\beq
\eta = \frac{b - a}{b + a}\,\,,
\label{Eq-04}\eeq
which is the relative difference of the half axis $a$ and $b$ of the cross
section perpendicular to the symmetry axis. Assuming that this parameter is the
same all across the shape, the profile function can be written in cylindrical 
coordinates in the form
\beq
 \varrho_s^2(z,\varphi) 
 = \rho_s^2(z) \, \frac{1-\eta^2}{1 + \eta^2 + 2 \eta \cos(2\varphi)}\,\,,
\label{Eq-05}\eeq
with $\rho_s^2(z)$ given by Eq.\ (\ref{Eq-02}). 
                                                    \\[ -2.0ex]

The above defined shape parametrization is rapidly converging, even for 
fissioning shapes, as demonstrated in Fig.\ 2 of Ref.\ \cite{SNP17}.
                                                                     \\[ -2.0ex]

A somehow odd feature of the Fourier coefficients $a^{}_n$ which
specify the shape, is their not necessarily transparent physical meaning. The 
Fourier coefficient $a_2$, for example, {\it decreases} with increasing
elongation. To cure this inconvenience, we have introduced the following four
new collective coordinates (see discussion in Ref.\ \cite{SNP17}) 
\begin{equation*}
 q_2 = \frac{a_2^{(0)}}{a_2} - \frac{a_2}{a_2^{(0)}} \,, \hspace{1.0cm}
 q_3 = a_3 \,, 
\end{equation*}
                 \\[ -5.0ex]
\begin{equation*}
 q_4 = a_4 + \sqrt{\left(\frac{q_2}{9}\right)^2 + \left(a_4^{(0)}\right)^2}\,\,,
\end{equation*}
                 \\[ -5.0ex]
\beq
 q_5 = a_5 - (q_2-2) \frac{a_3}{10} \,\,, 
\label{Eq-06}\eeq
                 \\[ -5.0ex]
\begin{equation*}
q_6 = a_6 - \sqrt{\left(\frac{q_2}{100}\right)^2 +\left(a_6^{(0)}\right)^2}\,\,,
\end{equation*}
where the $a_n^{(0)}$ are the values of the Fourier coefficients for the 
spherical shape. Note that the $q^{}_n$ coordinates have been defined in such a
way that they all vanish for a spherical shape. Their physical meaning is more
transparent and intuitive when compared to that of the $a^{}_n$ coefficients. In
the remainder of this work, we are therefore going to discuss the deformation
properties of nuclei in a 4D deformation space made of the collective
coordinates ($\eta$, $q_2$, $q_3$, $q_4$) which are directly related to
non-axiality, elongation, octupole, and hexadecapole (neck-thickness)
deformation. An alternative, but completely equivalent parametrization, based on
an expansion of the deviation of the nuclear shape from a spheroid is proposed
in Ref.~\cite{PNB17}.
                                \\[ +2.0ex]

\begin{figure}[!hbt]
\includegraphics[width=8.6cm]{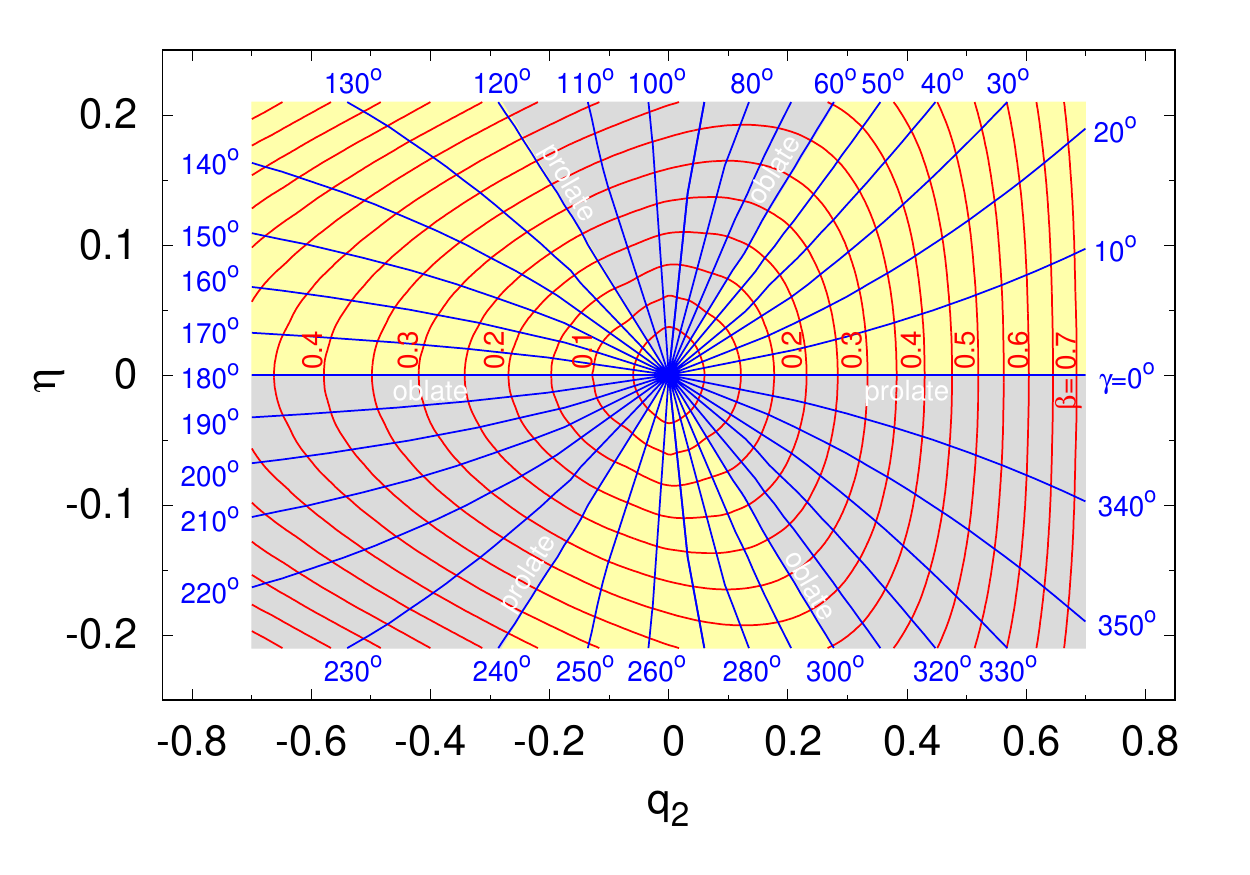}
\vspace{-0.8cm}
\caption{(Color online) Visualization of the relation between the ($\beta$,
$\gamma$) and the ($q^{}_2, \eta$) deformation coordinates.} 
\label{fig02}\end{figure}

The connection between the here-proposed ($q^{}_2, \eta$) coordinates and the
generally used ($\beta, \gamma$) deformation parameters \cite{Bo52} is
certainly worth to be discussed, in particular regarding the 60$^{\circ}$
symmetry of the latter. Let us recall that the ($\beta, \gamma$) variables are
defined as
\beq
 \beta = \frac{1}{X} \sqrt{Q^2_{20} + Q^2_{22}} 
  \hspace{0.6cm} \mbox{and} \hspace{0.6cm}
 \gamma = \arctan \left( \frac{Q^{}_{22}}{Q^{}_{20}} \right)
\label{Eq-07}\eeq
                 \\[ -1.0ex]
where $Q^2_{20}$ and $Q^2_{22}$ are the components of the mass quadrupole 
tensor
$$
 Q^{}_{20} = \langle 2 z^2 - r^2 \rangle \, , \hspace{0.8cm}
 Q^{}_{22} = \langle y^2 - x^2 \rangle \, ,
$$
where $X = 3 \, r_0^2 \, A^{5/3} / \sqrt{5 \pi} $ with the radius constant $r_0$.
                                                                     \\[ -2.0ex]

The connection between the ($\beta, \gamma$) and the ($q^{}_2, \eta$)
coordinates is shown in Fig.~\ref{fig02}, where the above mentioned 60$^{\circ}$
symmetry appears. This figure gives us the occasion to draw the attention of 
the reader to an inconvenience of both of these parametrizations at small 
deformation, and which calls for some cautious when interpreting the ($\eta$, 
$q_2$, $q_3$, $q_4$) deformation-energy landscapes. Due to the aforementioned 
symmetry, care has to be taken to avoid any ``double-counting'' of shapes. Let 
us take the example of an axially symmetric oblate shape defined by 
($q_2 \!=\!\! -0.35, \, \eta \!=\! 0$) corresponding to a deformation of
($\beta \!=\! 0.25,\, \gamma \!=\! 180^{\circ}$). From Fig.\ \ref{fig02} one concludes
that the same shape is realized by ($q_2 \!\approx\! +0.17, \, \eta \!\approx\!
0.13$), equivalent to ($\beta \!=\! 0.25,\, \gamma \!=\! 60^{\circ}$). A similar
correspondence can be established when considering prolate instead of oblate
shapes. For example, the configuration ($q_2 \!=\! 0.32, \, \eta \!=\! 0$) is
defined equivalently by ($\beta \!=\! 0.30,\, \gamma \!=\! 0^{\circ}$), and due
to the 60$^{\circ}$ symmetry, the latter defines the same shape as ($q_2 \approx
-0.19, \, \eta \approx 0.13$) equivalent to ($\beta \!=\! 0.25,\, \gamma
\!=\!120^{\circ}$). We emphasize that, such a strict comparison can only be 
carried out for spheroidal-deformed shapes. When higher order multipolarities 
come into play, and importantly, at larger deformation, this picture is 
partially distorted.
                                                                     \\[ -2.0ex]

Before closing the discussion on the relation between these two shape
parametrizations, we notice that a constant value of $\beta$ does {\it not}
correspond to a constant elongation of the shape, as can be seen from Fig.\
\ref{fig02}. That is why we believe that, for investigating the possibility of
triaxial shapes, the ($q^{}_2, \eta$) deformation space is better suited than
the traditional ($\beta, \gamma$) space.

\begin{figure*}[!h]
\begin{center}
\includegraphics[width=18.0cm]{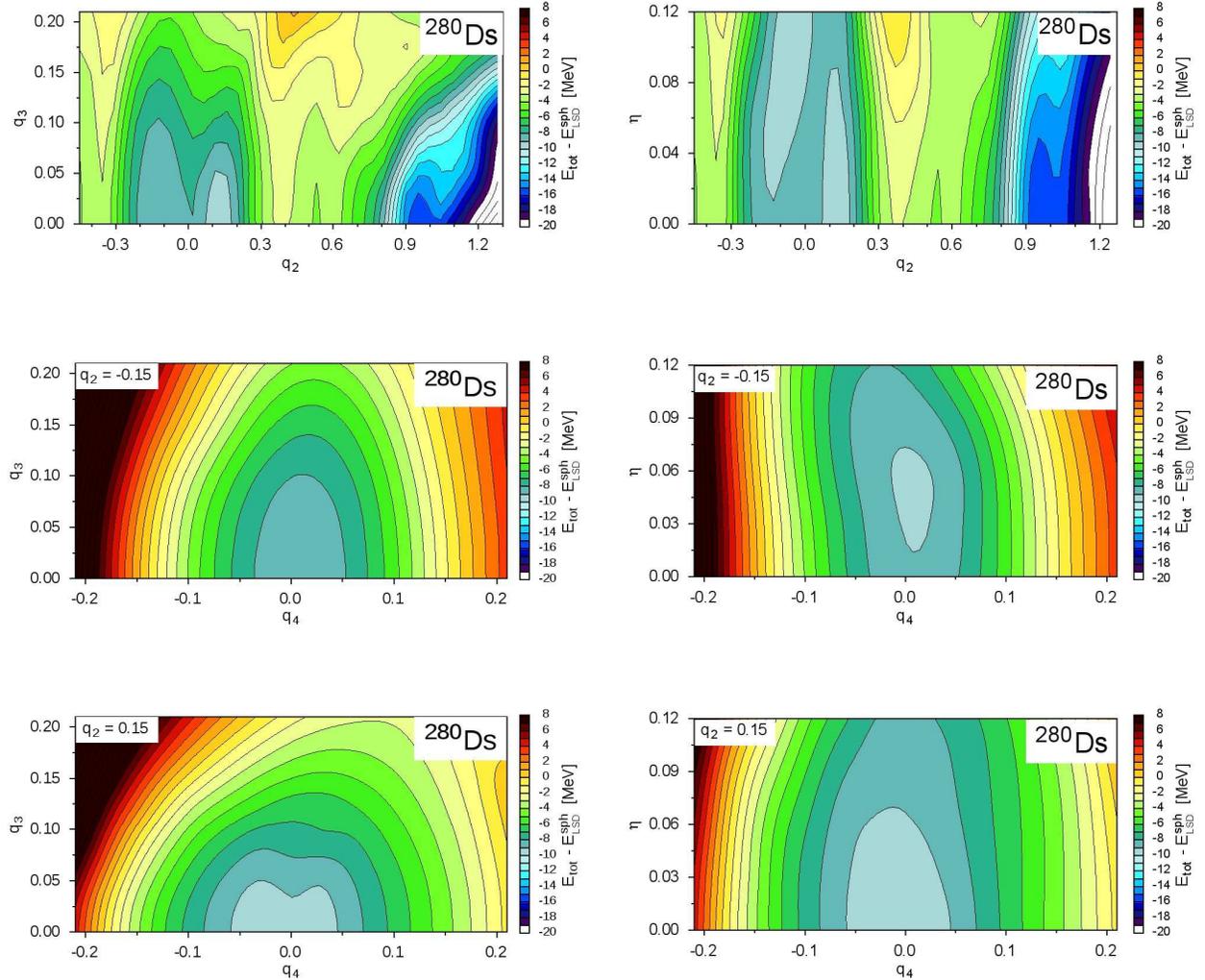}
\vspace{-0.3cm}
\caption{(Color online) Deformation energy of $^{280}$Ds on the $(q_2, q_3)$
plane for $\eta = 0$ minimized with respect to $q_4$ (top left) and on the
$(q_2,\eta)$ plane minimized with respect to $q_3$ and $q_4$ (top right).
The $(q_4,\ q_3)$ and $(q_4,\ \eta)$ cross sections around the oblate and
prolate minima are shown in the middle and bottom row, respectively.} 
\end{center}
\label{fig07}\end{figure*}
%

As an illustration of the symmetry property of both these shape 
parametrizations, and the caution to be applied in the analysis of 
deformation-energy landscapes (calculated as described in the next subsection),
let us consider the case of two deformations in the nuclei $^{280}$Ds and
$^{276}$Cn. Different cross sections of the 4D deformation space are presented
in Figs.~3 and 4. Each of these 2D landscapes has been realized by performing a
minimization with respect to the other two deformation coordinates, unless 
specified otherwise.
                                                                     \\[ -2.0ex]

The ($q^{}_2, q^{}_3$) deformation-energy map of $^{280}$Ds in Fig.~3 shows two
minima, one prolate ($q^{}_2 \!\approx\! +0.15$) and one oblate ($q^{}_2
\!\approx\! -0.15$). When investigating the ($q^{}_2, \eta$) cross section, one
notices, however, that the apparently oblate minimum corresponds, in fact, to a
triaxial solution ($\eta \!\approx\! 0.05$). When looking again at Fig.\ 2, one
concludes that a deformation $(q^{}_2 \approx -0.15, \eta \approx 0.05)$
characterizes the same shape as $(q^{}_2 \approx 0.1, \eta \approx 0)$ which is
nothing but our prolate shape. In addition, the energy landscape in $\eta$
direction is almost flat around the two minima. All this is a clear indication
that this ``oblate'' minimum is nothing but the {\it mirroring} of the (true)
prolate ground state, and does not correspond at all to a stationary point in
the deformation-energy landscape that has anything to do with a true oblate
deformation.
                                                                     \\[ -2.0ex]

\begin{figure*}[!htbp]
\begin{center}
\includegraphics[width=18.0cm]{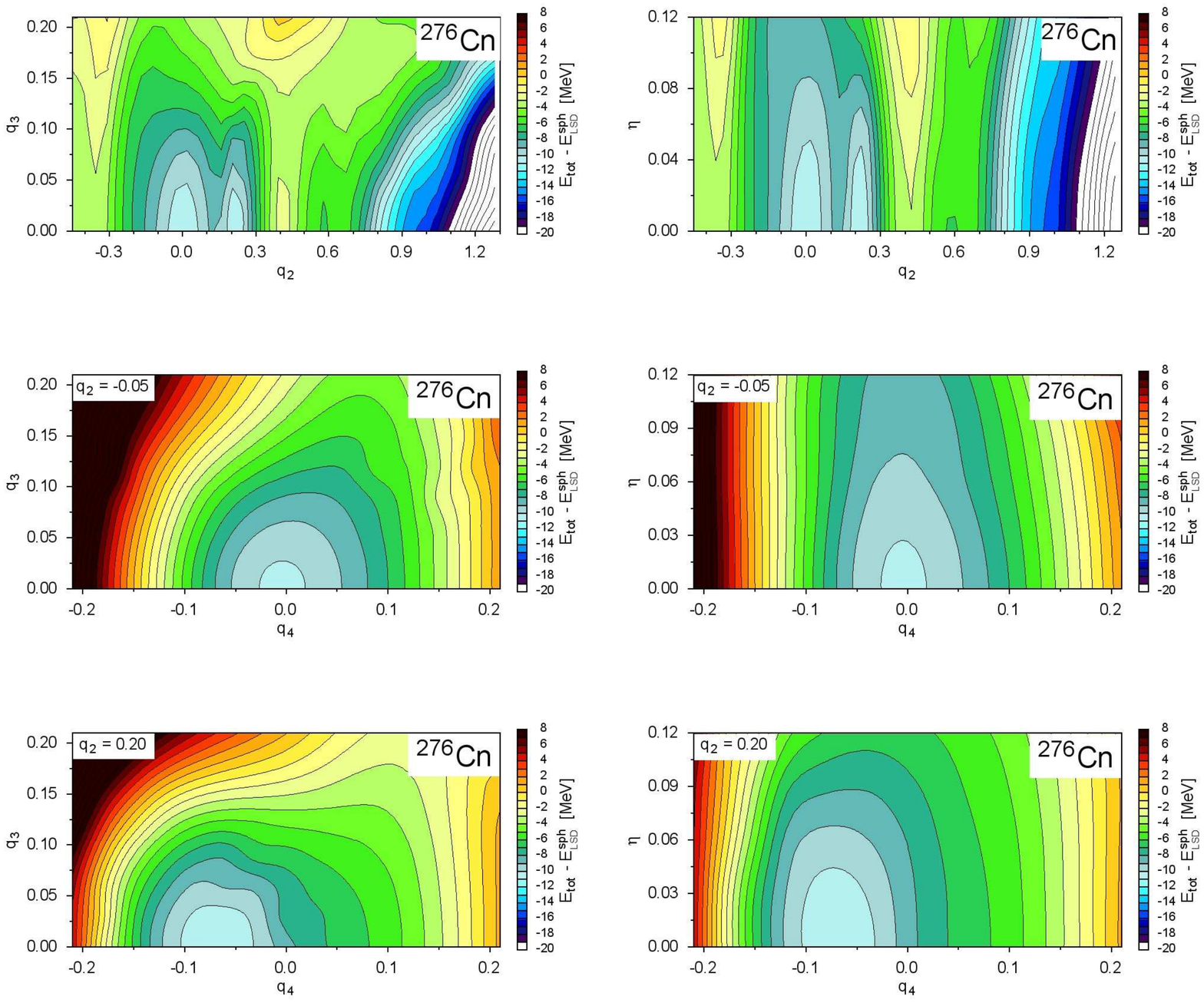}
\vspace{-0.3cm}
\caption{(Color online) Same as in Fig.~3 but for $^{276}$Cn.} 
\end{center}
\label{fig08}\end{figure*}

The situation is different for the case of the $^{276}$Cn nucleus illustrated 
in Fig.~4. Again two local minima are observed, one for a spherical shape
($q^{}_2 \!=\! 0$) and one for a prolate configuration ($q^{}_2 \!\approx\!
+0.23$). When looking at the ($q^{}_2, \eta$) cross section, both solutions
appear to be axially symmetric and are well separated in energy. This indicates
a true shape coexistence with both local minima being left-right symmetric
($q^{}_3 \!=\! 0$) (middle and bottom rows of Fig.\ 4, left) and having slightly
different hexadecapole deformation (middle and bottom rows of Fig.\ 4, right).\\


\subsection{The macroscopic-microscopic potential energy}\label{epot}

The potential energy of a nuclear system is calculated in our approach within
the macroscopic-microscopic model using the Lublin-Strasbourg Drop (LSD)
\cite{LSD} for the liquid-drop-type energy, including a curvature $A^{1/3}$ 
term in the leptodermous expansion and a deformation dependent congruence energy
term \cite{MS:1997}, which is well known to give a good description of nuclear 
ground-state masses and fission-barrier heights. The microscopic part is
determined by the Strutinsky shell-correction energies \cite{stru} and pairing
correlations \cite{MNM95} are derived in BCS theory with a seniority force and
an approximate particle-number projection \cite{GP86}. The single-particle
energies and wave functions that enter such an approach are obtained as
eigenvalues and eigenstates of a Yukawa-folded mean-field potential
\cite{YF,davies:1976} at given ($\eta$, $q_2$, $q_3$, $q_4$) deformation. More
details of the calculation are given in our previous work \cite{SNP17}.


\section{Results}\label{results}

Within the above-outlined theoretical framework, the 4D potential-energy
landscapes of 324 even-even nuclei with charge numbers in the range 
$92\le Z\le 126$ and isospins $40\le N-Z\le 74$
are evaluated. The considered grid consists of 32$\,$202 points in the ($\eta$,
$q_2$, $q_3$, $q_4$) space, with the following mesh:\\
\\
${}\hspace{0.5cm} \eta$ = 0 (0.03) 0.12 \\[0.7ex]
${}\hspace{0.5cm} q_2$ = -0.45 (0.05) 2.35 \\[0.7ex]
${}\hspace{0.5cm} q_3$ = 0 (0.03) 0.21 \\[0.7ex]
${}\hspace{0.5cm} q_4$ = -0.21 (0.03) 0.21 
                                  \\[ -1.0ex]

The calculated landscapes are analyzed, looking for ground-state (and possible
isomeric) equilibrium deformations, the corresponding $\alpha$-decay
properties, fission barrier heights and the most probable fission paths in the
region across Fm. The results of these investigations are reported in the
following subsections.

\subsection{Equilibrium configurations}\label{equilib}

For each isotope, the ground-state equilibrium energy and shape were determined
using the gradient oriented bisection method \cite{KP017} and the Gauss-Hermite
approximation \cite{PO16}. A so-determined equilibrium configuration is
characterized by the collective coordinates ($\eta^{eq}$, $q_2^{eq}$,
$q_3^{eq}$, $q_4^{eq}$) the values of which are displayed as a function of 
$Z$ and $N\!-\!Z$ in Fig.\ 5.

\begin{figure}[!hbt]
\begin{center}
\includegraphics[width=8.4cm]{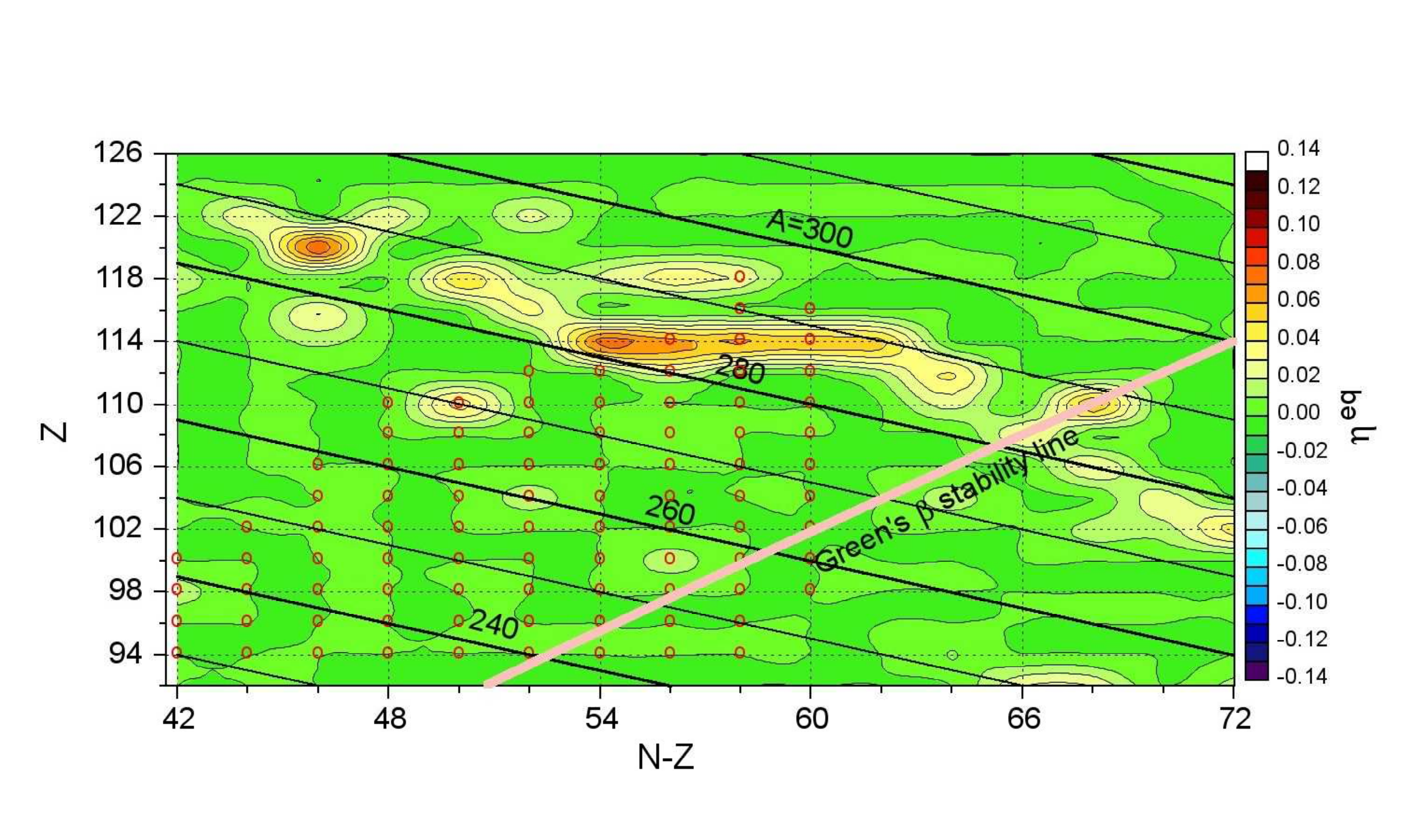}\\[-2ex]
\includegraphics[width=8.4cm]{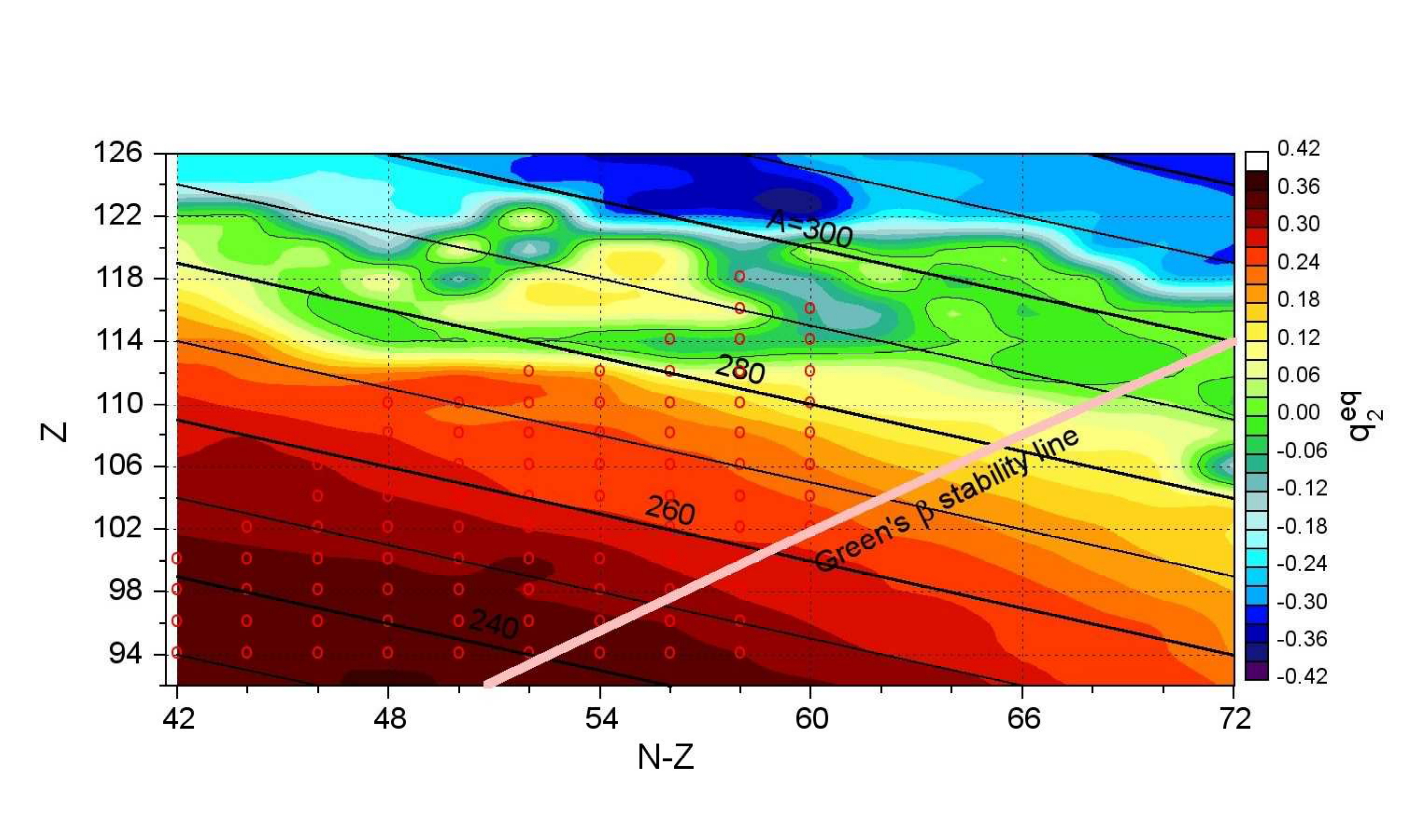}\\[-2ex]
\includegraphics[width=8.4cm]{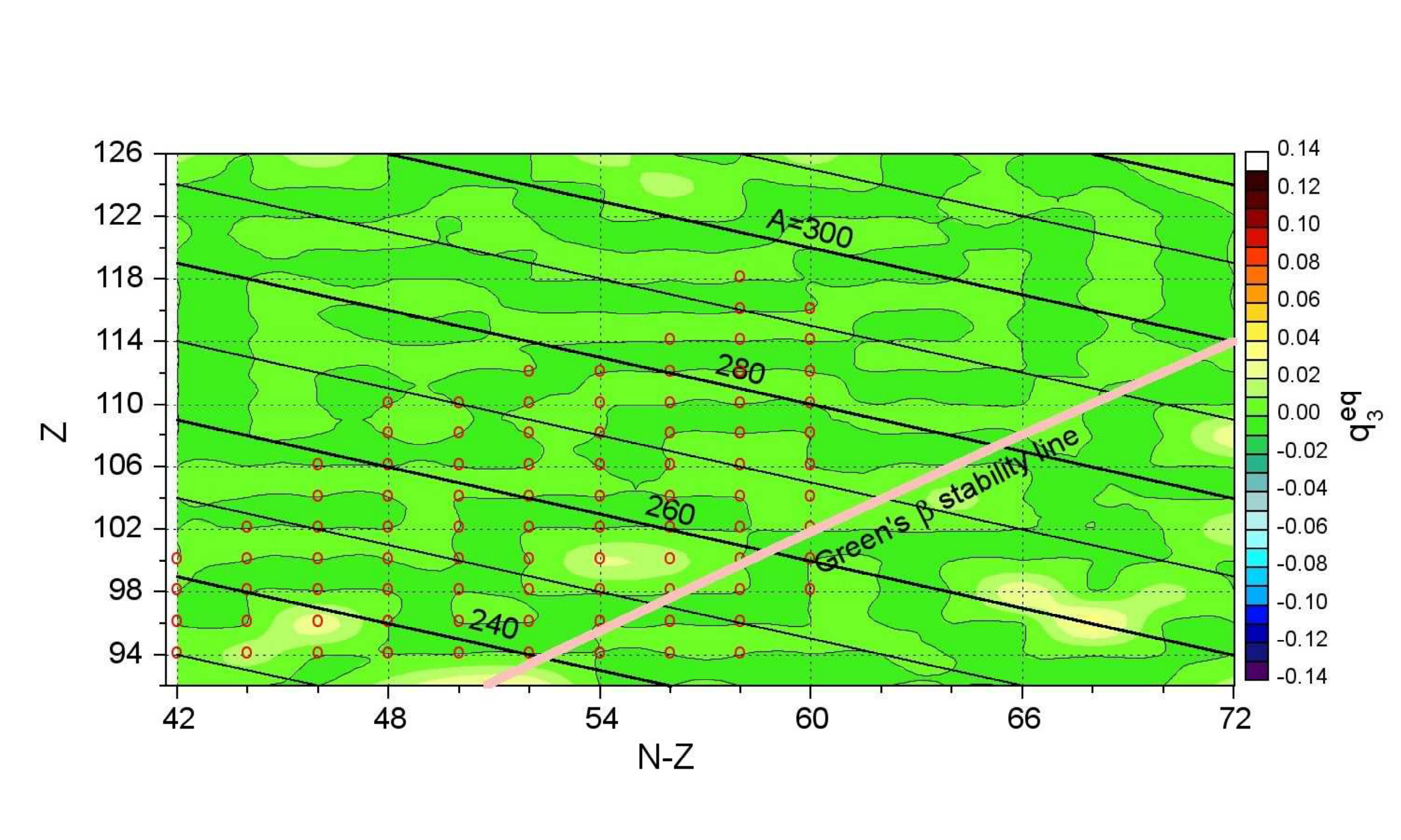}\\[-2ex]
\includegraphics[width=8.4cm]{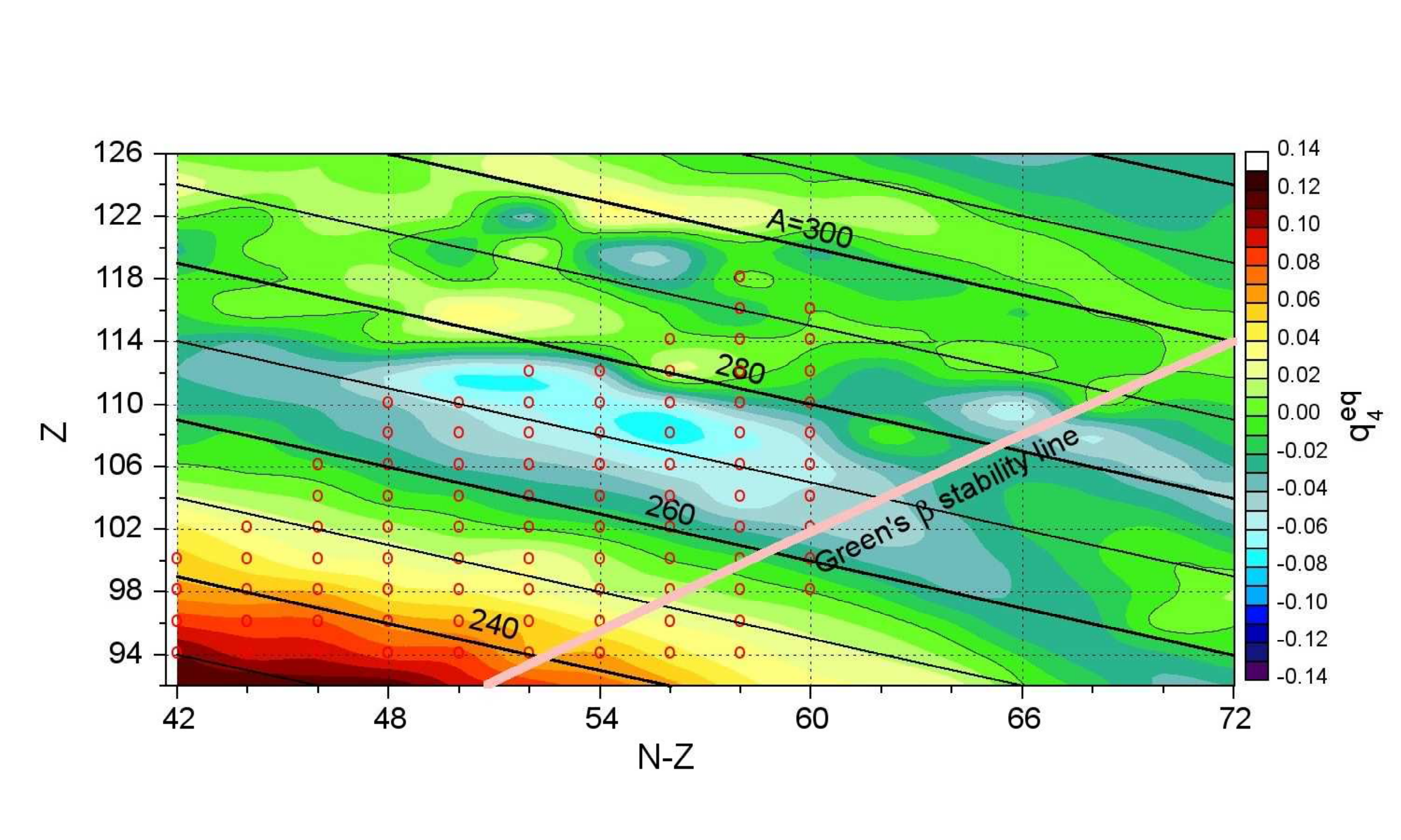}
\vspace{-0.2cm}
\caption{(Color online) Values of the collective coordinates ($\eta^{eq}$,
$q_2^{eq}$, $q_3^{eq}$, $q_4^{eq}$) (from top to bottom) at equilibrium
deformation defined as the ground-state minimal-energy configuration on the
($N-Z$, $Z$) plane. The circles denote already discovered isotopes and the thick (pink) line shows the $\beta$-stable nuclei. The thin (black) lines correspond to the constant mass number A.} 
\end{center}
\label{fig033}\end{figure}
The top part of Fig.\ 5 shows that most of the systems considered in this
work are axially symmetric ($\eta^{\rm eq} \approx 0$). A few exceptions are
noteworthy to discuss, though. First, one observes two somehow isolated cases:
the element Ds ($Z \!=\! 110$) which, in particular for $A \approx 270$, seems 
to be triaxial, as well as $^{292}$Og ($Z \!=\! 118, N \!=\! 174$). 
At first sight, the nuclear system $(Z \!=\!120,\,N \!=\! 166)$ 
seems to possess a quite strong non-axiality. Having, however, a 
closer look at its quadrupole parameter $q_2$ (second panel in Fig.\ 5), one 
notices that this parameter turns out to be almost vanishing, and, as we have 
explained before, near the spherical shape, the non-axiality degree of 
freedom loses its meaning.
In addition one observes, for $102 \le Z \le 110$, a diagonal downward-sloping
band running over the isospin interval $64\le N-Z\le 72$. The corresponding
isotopes (from $^{276}$No to $^{284}$Ds) are predicted to possess a slight
triaxiality, and/or to be soft with respect to triaxial deformations. It is
interesting to notice that all nuclei in this band have a neutron number $N$
around 174. As we will see in the next subsection, these nuclei are
prolate-deformed. Hence, the present model predicts a triaxial "window" across
$N \approx 174$ for elements from No to Ds. We nevertheless emphasize that a
value of $\eta \approx 0.07$ corresponds to a non-axial deformation where the
longer of the two half axis is only 15\% larger than the shorter one. That is to
say that the observed effect is small. Please also note that for Fl ($Z = 114$)
isotopes with isospin $N-Z$ values between 54 and 62, for which a non-zero
$\eta^{\rm eq}$ value is deduced from the top part of Fig.\ 5, have a
practically vanishing $q_2^{\rm eq}$ value (second panel), and are therefore
close to a spherical shape where the non-axiality parameters $\eta$ and
$\gamma$ loose their meaning. The observation about a dominating axial symmetry
in the region, and the occurrence of a weakly triaxial ``window'' around $N
\approx 174$ for No to Ds is consistent with predictions by other models, like
e.g.\ the self-consistent approaches of Ref.\ \cite{cwiok2005,bruyeres17}.
                                  \\[ -2.5ex]
              
Equilibrium quadrupole-type deformations are investigated in the second panel of
Fig.\ 5. Transuranic elements with masses below $A \!\approx\! 280$ are observed
to possess a prolate ground-state deformation, while beyond $A \!\approx\! 300$
an oblate configuration is predicted, especially in the vicinity of the nucleus
$^{304}$122. It is interesting to note that, up to around $Z = 114$ the
magnitude of the prolate elongation steadily decreases for all elements with
increasing $N$ (or equivalently $N-Z$). For heavier elements the dependence on
$N$ is more erratic. These different trends depending on $Z$ are very likely to
be connected to the evolution of shell corrections with $Z$ and $N$, as further
discussed below. A similar trend for quadrupole deformation across the region
was anticipated both with macroscopic-microscopic and self-consistent methods
(see e.g. Refs.~\cite{cwiok2005, cwiok1996, moller2012}).
                                                                     \\[ -2.0ex]

Left-right asymmetric shapes are investigated in the third panel of Fig.\ 5. One
concludes that the ground-state is expected to be essentially left-right
symmetric ($q_3^{\rm eq} \approx 0$) for all nuclei in the region. 
                                                                     \\[ -2.0ex]

Finally, the last panel of Fig.\ 5 suggests that, while the lighter actinides
with $A \approx 230$ have a substantial hexadecapole ground-state deformation
($q_4^{\rm eq} \!\approx\! +0.10$), corresponding to a diamond-like (rugby-ball
type) configuration, those in the region between Hs and Cn with $A \!\approx\!
270$ are predicted to have strong negative hexadecapole deformation ($q_4^{\rm
eq} \!\approx\! -0.08$), looking a bit like a rounded-off rod. Again, one notes
a rather steady evolution with $N$ for elements up to $Z \approx 108$, and
presumably structural effects for higher $Z$'s.
                                                                     \\[ -2.0ex]

Since the macroscopic energy has no minimum for SHE isotopes, the equilibrium
configuration is governed by shell effects. The microscopic contribution defined
as the total potential energy at equilibrium relative to the macroscopic energy
at spherical shape, $\delta E^{eq}_{\rm mic} = E^{eq}_{\rm tot} - E^{sph}_{\rm
LSD}$, is shown in Fig.\ 6, as in Fig.\ 5, on the ($N-Z$, $Z$) plane. Two
regions of strong microscopic effects are visible. The island centered around 
$Z = 108$, $N-Z = 54$ is driven by the deformed $N = 162$ shell; large prolate
deformation can, indeed, be deduced from the second panel of Fig.\ 5. The
evidence for stabilization through deformation in the $^{270}$Hs region was
experimentally confirmed \cite{zzz}. Our calculations predict even stronger
microscopic effects in a band running from $Z \approx 114$ to $Z \approx 118$
and with $48 \le N \!-\!Z \le 66$. The effect is largest at $Z = 116$ for the
$^{286}$Lv and $^{290}$Lv isotopes, corresponding to $N$ = 170 and 174. Figure 5
shows that nuclei in this band are characterized by a weak prolate deformation.
The present model therefore predicts that the next ``magic configuration'' (not
yet reached by the experiment) would be slightly prolate-deformed and located
at $Z = 114 - 116$ and $N = 170 - 174$. A separate analysis of the proton and
neutron microscopic corrections shows that this stabilization is mainly driven
by the neutrons in our calculations. No evidence for spherical magicity at 
$Z = 114$ and $N = 184$ is apparently seen in our results, contrary to what was
anticipated in other macroscopic-microscopic models
\cite{kowal2014,MOLLER09,MOLLER15}.
                                  \\[ -2.7ex]

\begin{figure}[!hbt]
\includegraphics[width=\columnwidth,height=6.cm]{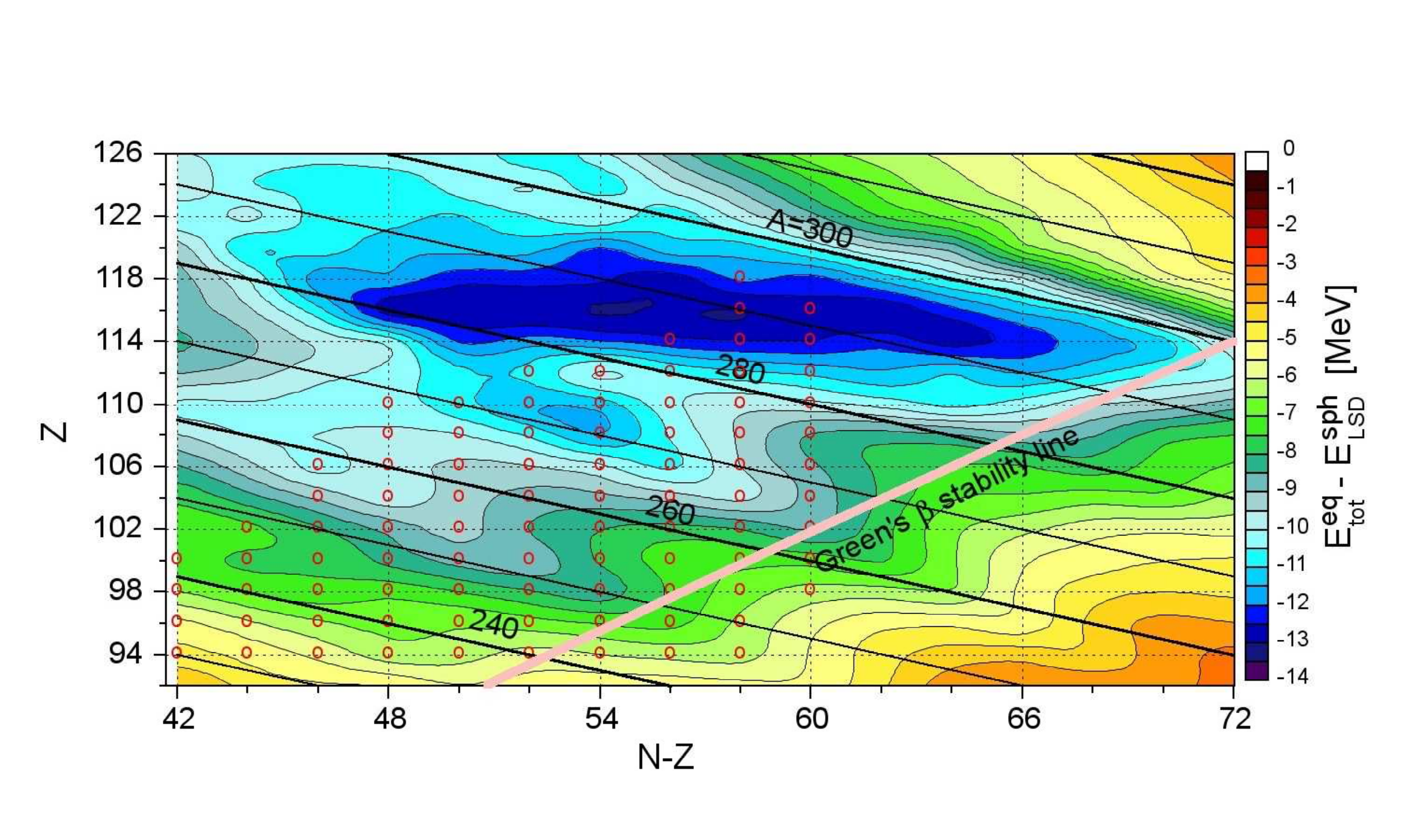}
\caption{(Color online) Ground-state microscopic contribution to the potential
energy for the nuclei of Fig.\ 5.} 
\label{fig04}
\end{figure}

\begin{figure}[!hbt]
\includegraphics[width=8.0cm]{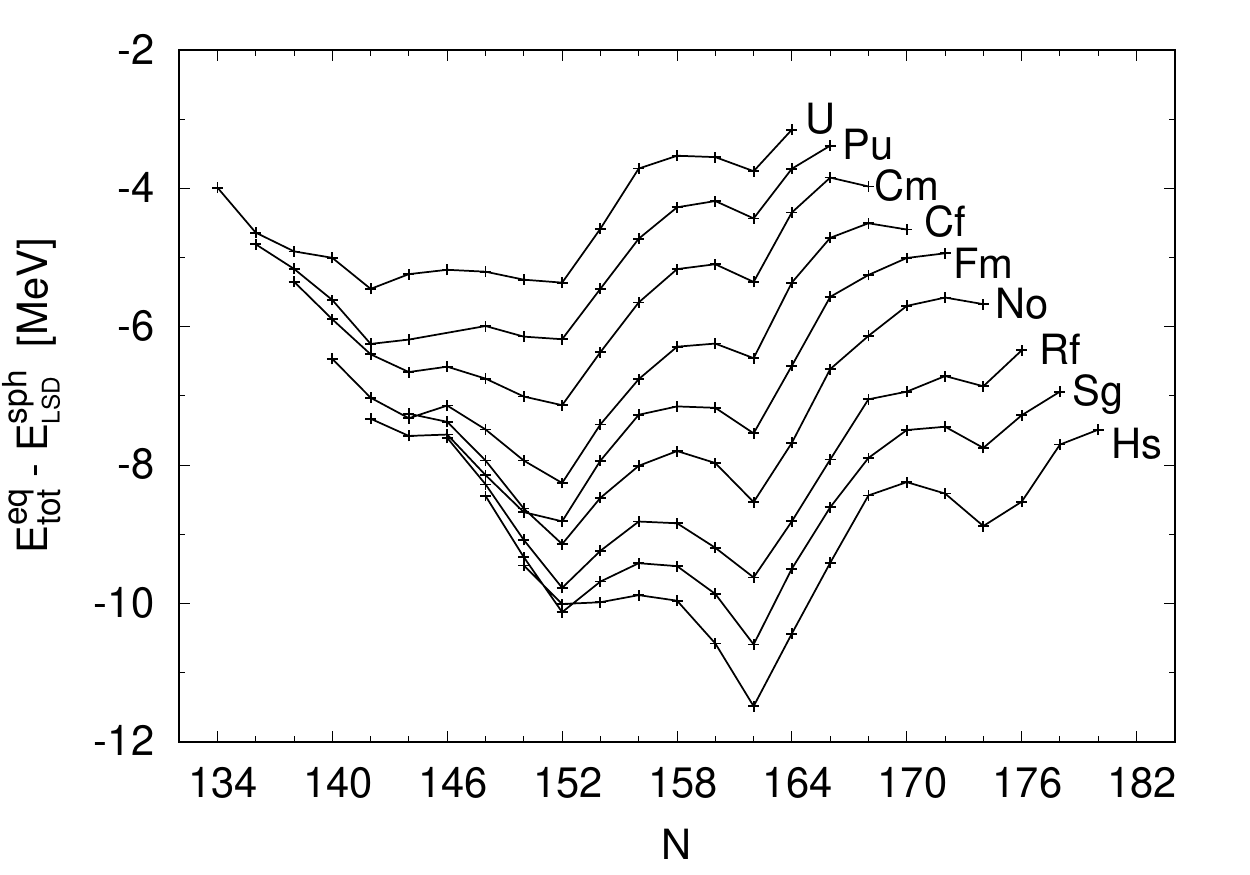}
\includegraphics[width=8.0cm]{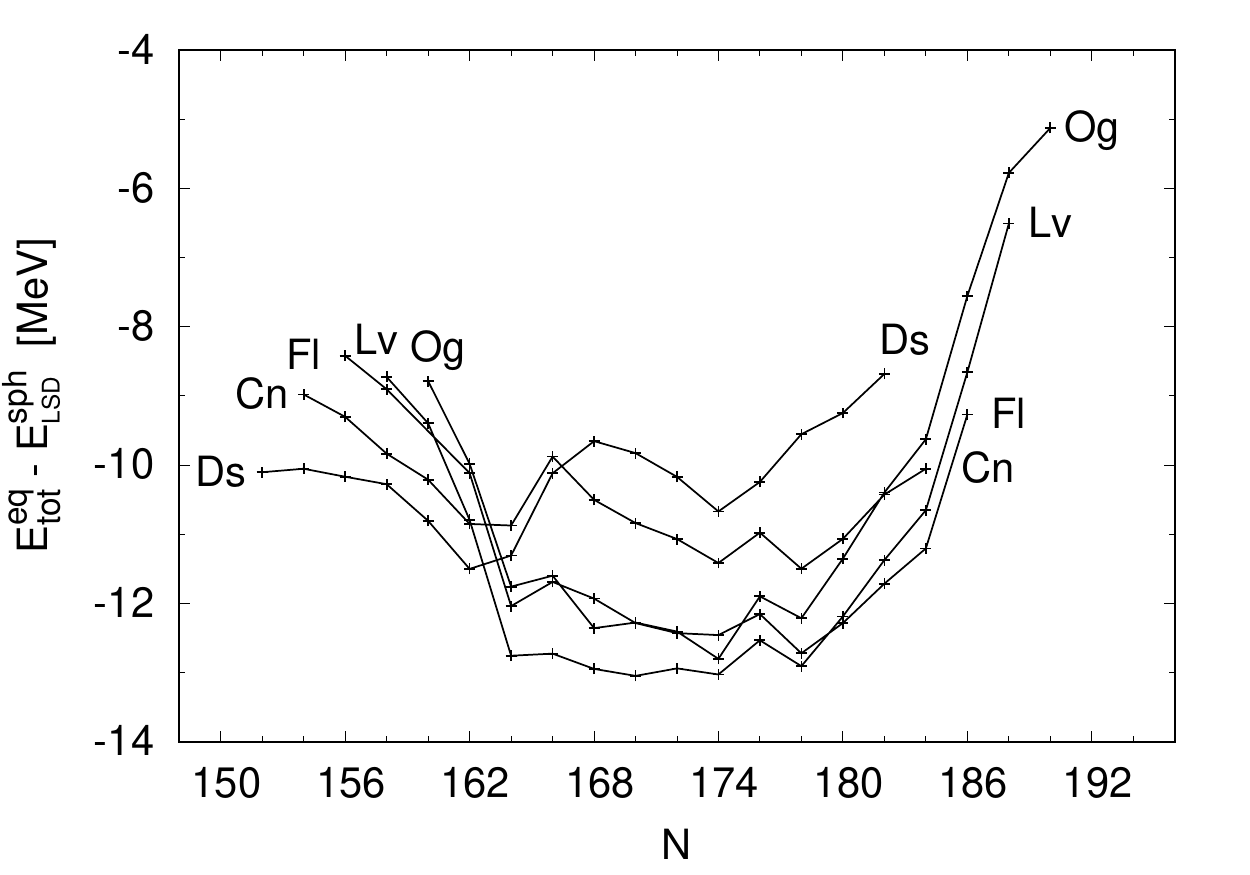}
\includegraphics[width=8.0cm]{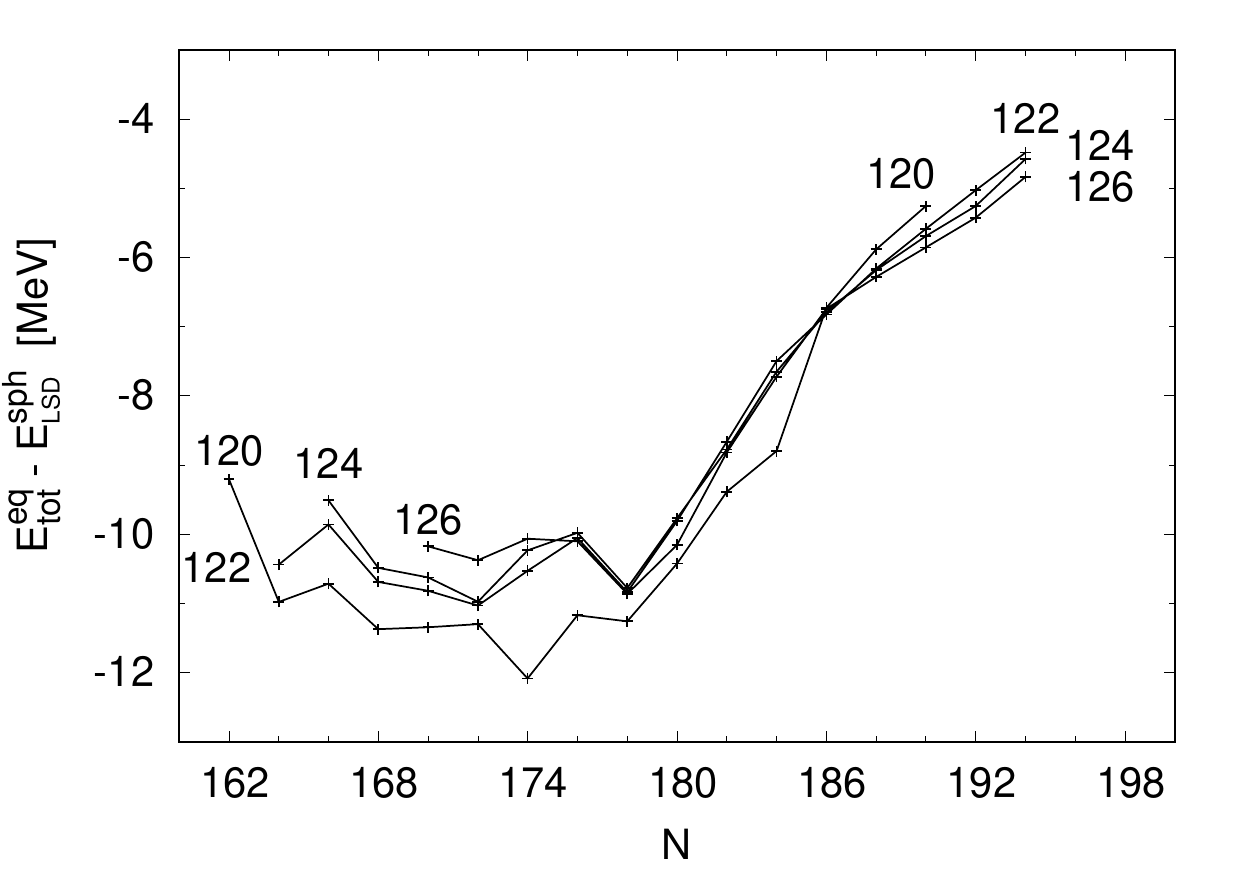}
\vspace{-0.3cm}
\caption{(Color online) Ground-state microscopic contribution to the potential
energy along isotopic chains between U and Hs (top), between Ds and Og
(middle), and between elements $Z = 120$ and $Z = 126$ (bottom).} 
\label{fig05}\end{figure}

While Fig.\ \ref{fig04} gives a condensed overview for the nuclei in the whole
region, a more detailed quantitative visualization is proposed in Fig.\
\ref{fig05}. There, the evolution of the quantal corrections to the ground-state
energy is shown separately for each isotopic chain, being displayed for all $Z$
values included in the 2D overview of Fig.\ \ref{fig04} as function of the
neutron number. The shell-stabilized configurations discussed above appear as
local, more or less abrupt, dips. The deformed shell effect at $N = 162$ that
develops from U towards Hs, and vanishes beyond Cn, is clearly seen. Although of
weaker magnitude, a similar shell effect becomes visible at $N = 152$ for
elements between Cm and Sg, which is also corroborated by experimental
observation \cite{zzz}. The $N = 152$ deformed region is located in Fig.\
\ref{fig04} ``southwest'' of the $N = 162$ island. Finally, the new predicted
``SHE magic shell closure'' discussed above first appears as a dip at $N = 174$,
starting from say Ds. For heavier elements, the deformed $N = 162$ and $N = 174$
shells merge, producing the wide band of strong stabilization discussed
previously from $Z \approx 114$ to $Z \approx 118$ and with $168 \le N \le 178$.
                                                                     \\[ -2.0ex]

\begin{figure*}[!h]
\begin{center}
\includegraphics[width=18cm]{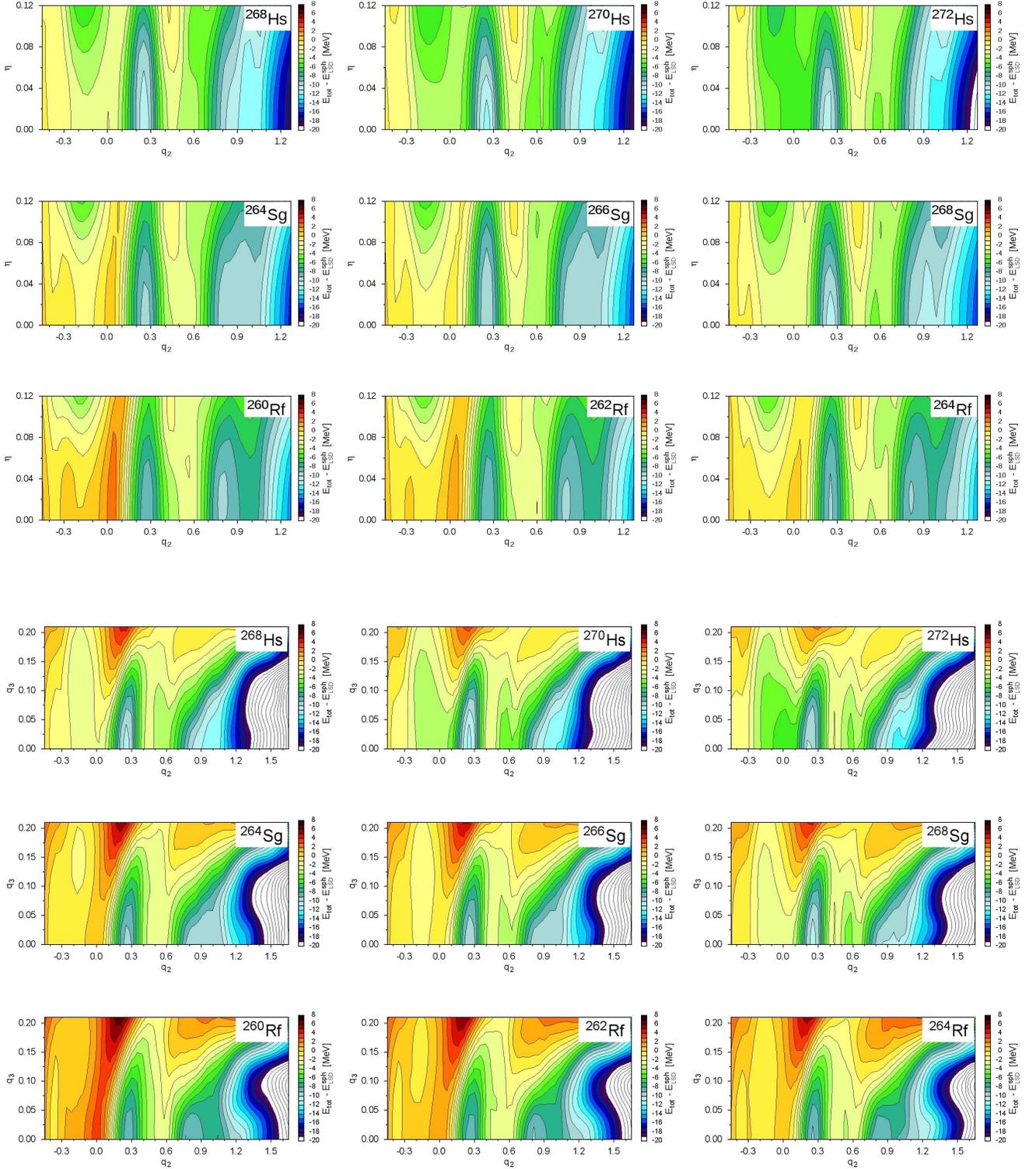}
\vspace{-2cm}
\caption{(Color online) Deformation energy in the ($q_2, \eta$) (upper half) and
($q_2, \, q_3$) (lower half) sub-spaces for three isotopes of Hs (first row), Sg
(second row) and Rf (third row). Each column corresponds to a specific $N - Z$
value: 52 (left), 54 (middle) and 56 (right).}
\end{center}
\label{fig06}\end{figure*}

The presentation adopted in Figs.\ 5 and 6 confers a fast overview of the
ground-state properties across the entire studied region. It is the result of
the analysis of the theoretical 4D potential-energy landscapes. To illustrate
the richness of these landscapes in more detail, their progressive evolution
with proton and neutron numbers, and the possible occurrence of metastable,
isomeric states, 2D cross sections of the 4D deformation space are presented in
Fig.\ 8. The deformation energy, defined as the total potential energy relative
to the spherical liquid-drop energy, is displayed for three isotopes of the Rf,
Sg and Hs elements, in the ($q_2$, $\eta$) (top part of the figure) and ($q_2$,
$q_3$) (bottom part of the figure) subspaces. Each of the 2D surfaces is
obtained, as already mentioned above, after minimization with respect to the
two remaining collective variables. The selected isotopes for a given $Z$ are
characterized by the same values of the isospin $N-Z = 52$, 54 and 56. The
corresponding $N$ values vary between 156 and 164. From the ($q_2$, $q_3$)
landscapes, all these nuclei are found to have a prolate ground-state
deformation with $q^{}_2 \approx 0.28$, and to be left-right symmetric ($q_3 =
0$). The analysis of the ($q_2$, $\eta$) landscapes further confirms the axial
symmetry of these isotopes.
                                  \\[ -2.7ex]

It is interesting to observe in Fig.\ 8 (not visible in Fig.\ 5), that there is
a left-right symmetric shape-isomeric state which appears for the Rf isotopes at
$q^{}_2 \approx 0.8$, and, though less pronounced, for Sg and Hs at $q^{}_2
\approx 0.6$. The occurrence of the isomeric local minimum is seen to strongly
depend on {\it both} $N$ and $Z$. Further, for the selected isotopes having the
same $N - Z$, the observed evolution implies that the appearance of this state
is not governed by $N - Z$ either. Altogether the appearance of this isomeric
state suggests that it originates from the subtle interplay of neutron and
proton microscopic effects at large deformation.

\subsection{Alpha-decay properties}\label{alphadecay}

Alpha radioactivity is the dominant decay mode in the VHE and SHE region. Most
of the currently known heaviest nuclei have actually been identified due to
their connection via $\alpha$-decay (see \cite{OGALL} and Refs.~therein) to
their previously known daughter nucleus. The availability of reliable
$Q_{\alpha}$ predictions is therefore crucial for an assignment of new elements.
The energy released when the nucleus emits an $\alpha$ particle is directly
related to the nuclear masses. Hence, $Q_{\alpha}$ values are also an indirect
way to test nuclear-mass models. The macroscopic-microscopic model used in the
present work has shown to provide a particularly good description of nuclear
masses \cite{LSD}. Reliable predictions for $Q_{\alpha}$ values are therefore
anticipated. 
                                                                     \\[ -2.0ex]
 
\begin{figure*}[!htbp]
\includegraphics[width=18.0cm]{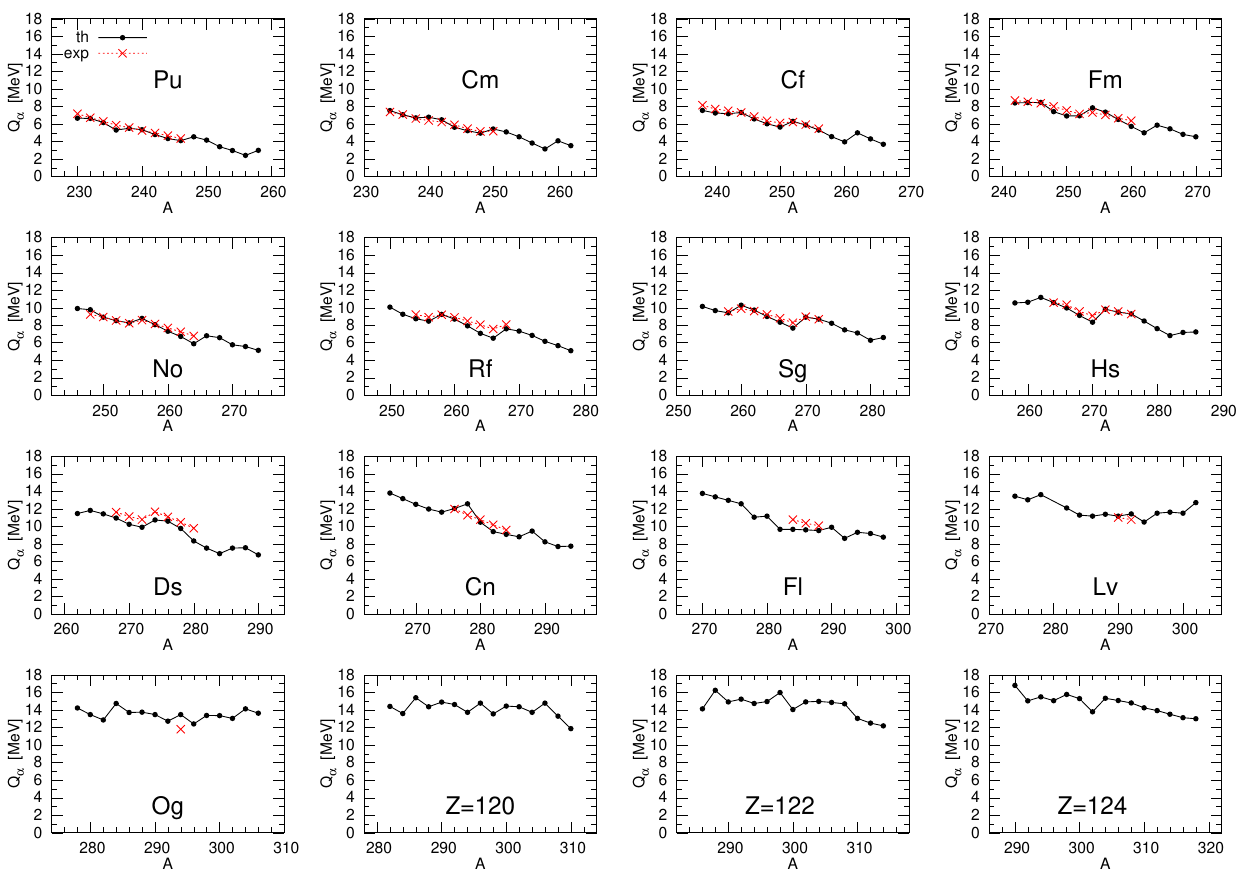}
\vspace{-0.4cm}
\caption{(Color online) Calculated $Q_\alpha$ energies for different isotopic
chains compared to the experimental data (red crosses) \cite{ccc} where available.} 
\label{fig11}\end{figure*}

The calculated $Q_\alpha$ are displayed in Fig. \ref{fig11} as function of the 
mass number for different isotopic chains from Pu up to element $Z = 124$. The
aforementioned $N = 162$ deformed shell appears as a dip at the corresponding
mass for elements between Fm and Ds. Similarly, a local minimum is predicted,
due to some kind of shell closure in the Sg, Hs, Ds and Cn isotopic chains, at
$N \approx 174$.  To explain the change of magic numbers from $N = 162$ to $N =
174$, please notice that it is precisely in this mass region, when going from 
$A \approx 270$ to $A \approx 290$, that the ground-state deformation changes 
from strongly prolate to close to spherical, as can be seen on the second panel
of Fig.\ 5. The weak microscopic effect at $N = 152$ appears as a very shallow
minimum beyond Cm up to Sg. One notices the absence of a clear and systematic
local minimum at $N = 184$, in contrast to the predictions of other models (see
e.g. Ref.~\cite{kowal2014}). A comparison of the results of our calculations
with the experiment \cite{ccc}, wherever available, is observed to be very good.
The apparently good description of nuclei, as function of $Z$, at and around $N
= 152$ and 162, is particularly noteworthy.
                                                                     \\[ -2.0ex]
 
\begin{figure*}[!htbp]
\includegraphics[width=18.0cm]{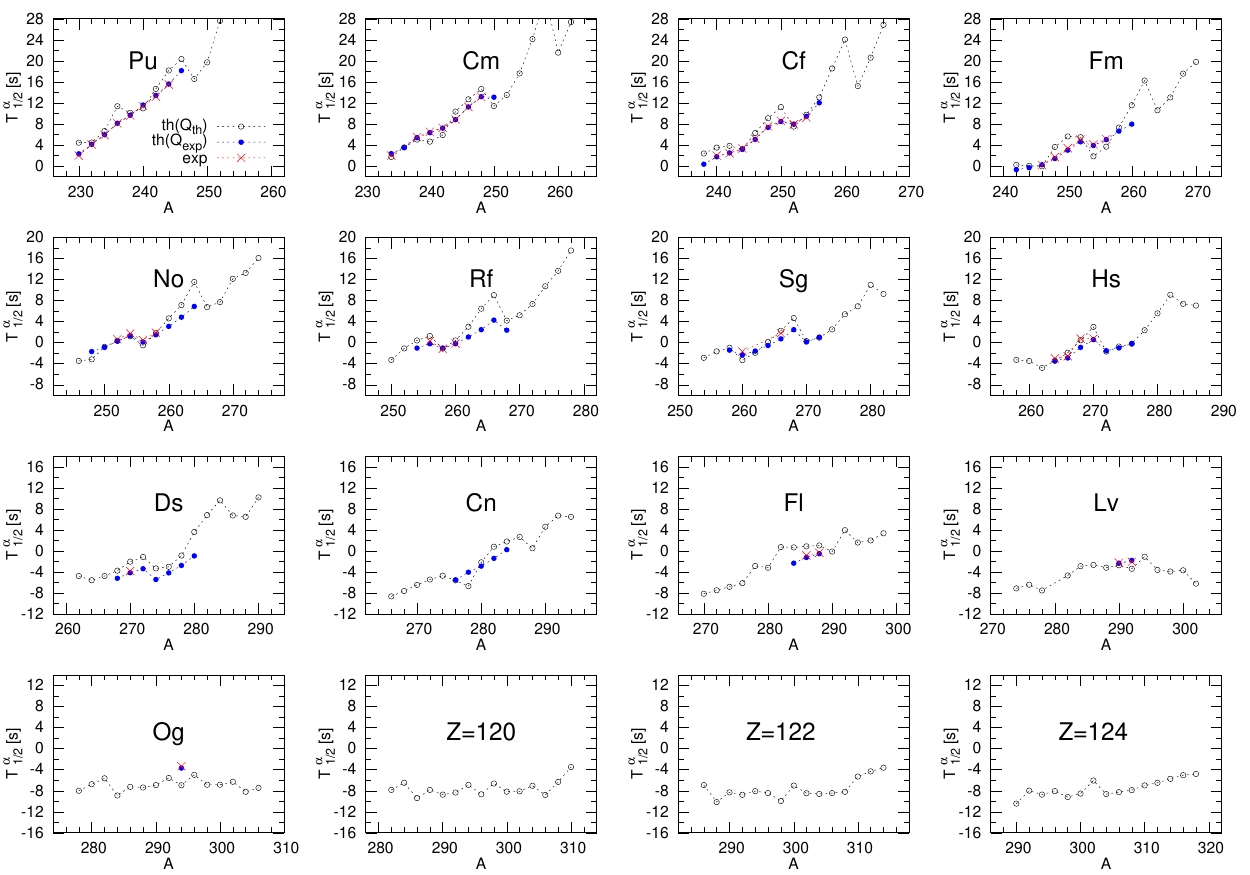}
\vspace{-0.4cm}
\caption{(Color online) Calculated ${\rm T}^{\alpha}_{1/2}$ $\alpha$-decay 
half-lives for different isotopic chains compared to the experimental data (red
crosses) \cite{ddd} where available. Two theoretical estimates given by the
open black and full blue circles are shown (see the text).} 
\label{fig12}\end{figure*}

Encouraged by the promising results of our model for $Q_{\alpha}$ energies, we 
compute the $\alpha$-decay half-life ${\rm T}^{\alpha}_{1/2}$ (for which
$Q_\alpha$ is the main ingredient) in the framework of the Gamow-type WKB
approach of Ref.\ \cite{ZWP13}, with no additional adjustment of any parameter.
The corresponding results are presented in Fig.\ \ref{fig12}. In addition to
the experimental half-lives, two types of theoretical estimates are displayed
there. Open black circles correspond to the $Q_\alpha$ energies produced by our
macroscopic-microscopic model, while full blue circles are obtained with the
experimental $Q_\alpha$ values where available. These last estimates are found
to agree almost perfectly with the experimental data \cite{ddd}, which, to our
understanding, is a clear indication of the value of our Gamow-type WKB
approach. The difference between the two theoretical curves demonstrates the
strong sensitivity of ${\rm T}^{\alpha}_{1/2}$ on the precise value of
$Q_{\alpha}$. Similar to the case of $\alpha$-decay energies, shell effects
lead to local mimina in the evolution of ${\rm T}^{\alpha}_{1/2}$ with mass
number $A$.

\subsection{Fission barriers}\label{barrier}

Besides $\alpha$-decay many VHE and SHE isotopes are characterized by a high
probability of decaying via spontaneous fission \cite{HESS2017}. Accurate
quantitative predictions of spontaneous fission properties remain a challenge
for theory, due to the poor knowledge of the many ingredients entering its
description (see \cite{robledo2013} and Refs.\ therein). Improving our
understanding of the process is important even for areas outside traditional
nuclear physics, like astrophysics \cite{GOR2013}.
                                                                     \\[ -2.0ex]

\begin{figure}[!h]
\includegraphics[width=\columnwidth,height=6.cm]{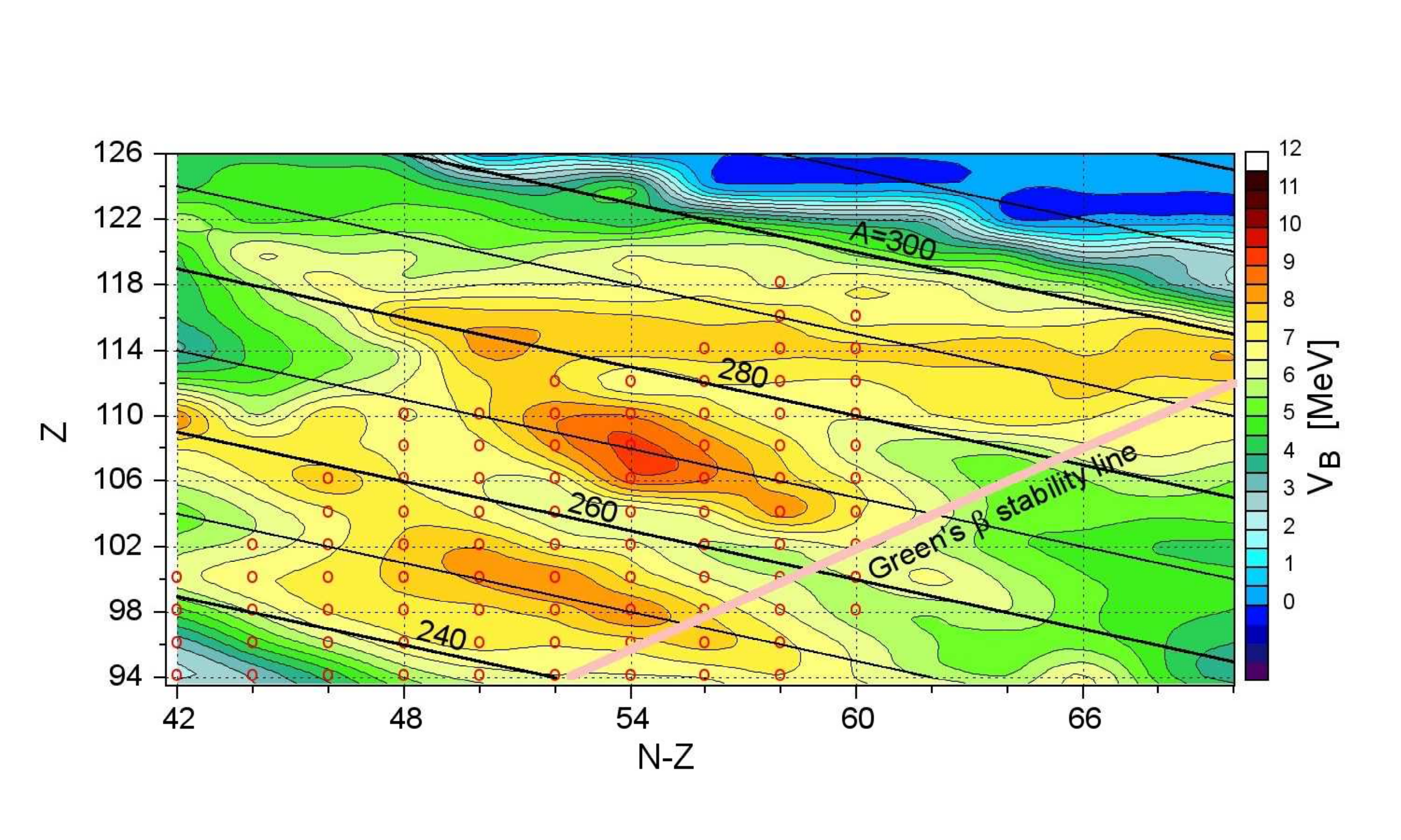}
\vspace{-0.3cm}
\caption{(Color online) Fission barrier height for the considered nuclei as 
function of $Z$ and $N-Z$.} 
\label{fig09}\end{figure}

To address the question of stability against spontaneous fission, the fission
barrier constitutes a crucial quantity. Its heights derived from the 4D
landscapes calculated in our approach are displayed in Fig.\ \ref{fig09} as 
function of $N-Z$ and $Z$. One notices a rather high fission barrier in the
region centered around $^{270}$Hs (V$_B \approx$ 9 MeV), but also around 
$^{252}$Fm (V$_B \approx$ 8 MeV). These islands of higher stability against
spontaneous fission are related to the $N =162$ and $N = 152$ shell effect,
respectively. Large barriers (V$_B \approx$ 8 MeV) are also predicted near 
$^{278}$Fl ($N=164$). These should lead to increased stability against fission
of the corresponding isotopes.
                                                                     \\[ -2.0ex]
 
Our estimates on fission barriers are consistent both qualitatively and
quantitatively with the predictions by the macroscopic-microscopic model of
Moller et al. \cite{MOLLER09,MOLLER15} in Hs and Fm isotopes. Similarly to
Moller, we again observe enhanced fission-barrier heights in a band $(Z \!=\!
114 \!-\! 118,\, N\!-\!Z = 48 \!-\! 66)$, even though the effect seems somehow
weaker in our approach as compared to his. Still slightly different predictions
are published in e.g. Refs.~\cite{kowal2017, baran2015, abusara2012}.
                                                                     \\[ -2.0ex]

Figure\ 11 displays our calculated barrier heights as function of mass number
$A$ for the same isotopic chains as in Figs. \ref{fig11} and \ref{fig12}.
Comparison with the (unfortunately sparse) experimental data \cite{fissbar1,
fissbar2,fissbar3, fissbar4} shows that our estimates are, indeed, very
reasonable. 
                                                                     \\[ -2.0ex]

In recent works \cite{PWZ15,rila2017} we have also applied the above-quoted simple
WKB approach to estimate the spontaneous fission half-lives. Comparison with
experiment between Th and Fl is presented in Fig.\ 10 of Ref.~\cite{PWZ15}.
The description by our model is impressively good, with deviations from the
experimental data which are on the average less than one order of magnitude what
is remarkable in the field (see e.g.\ discussion in Ref.~\cite{robledo2013}). 
                                                                     \\[ -2.0ex]

\begin{figure*}[!htbp]
\begin{center}
\includegraphics[width=17.0cm]{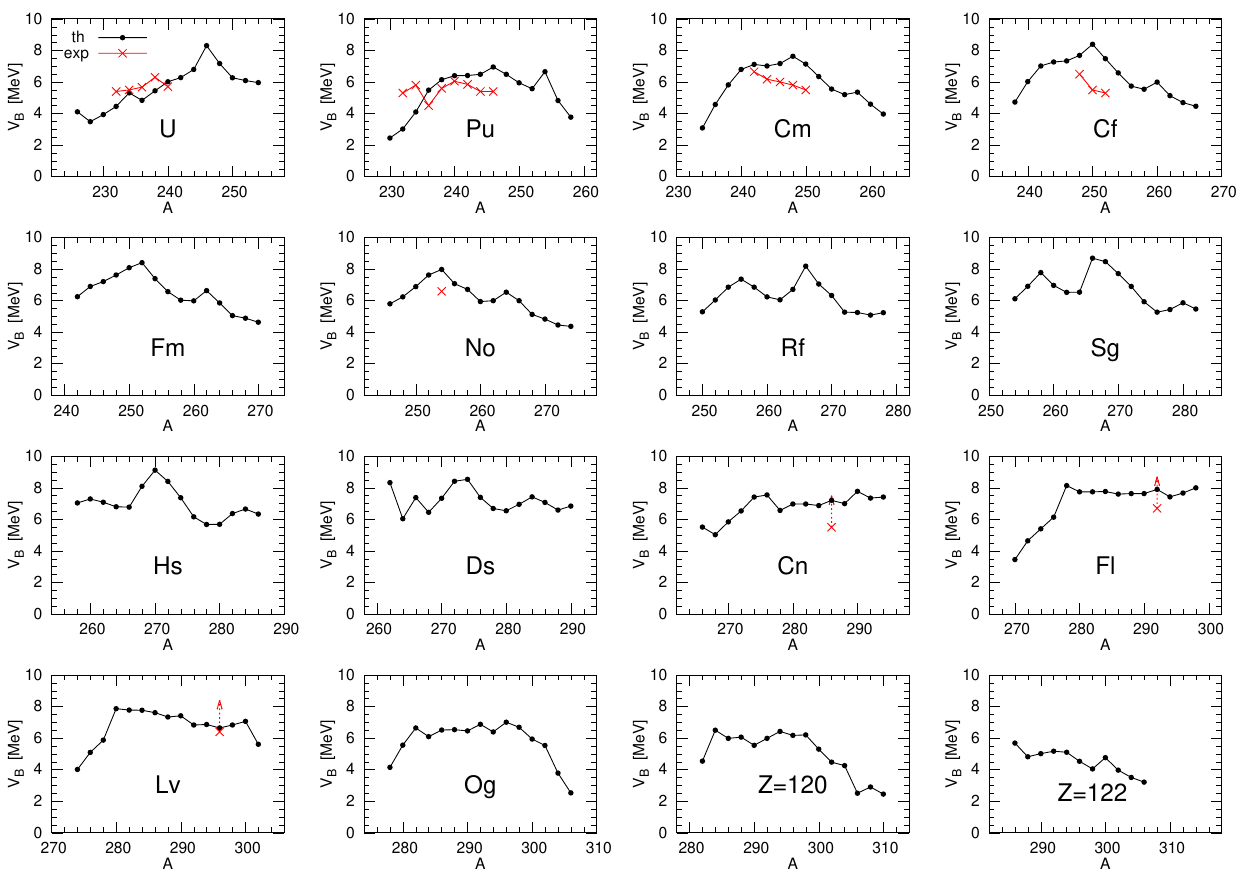}
\vspace{-0.4cm}
\caption{(Color online) Similar to Figs.\ 9 and 10 for calculated 
fission-barrier heights compared to the data \cite{fissbar1,
fissbar2,fissbar3, fissbar4}.} 
\end{center}
\label{fig10}\end{figure*}
\begin{figure*}[!hbtp]
\includegraphics[width=6.5cm]{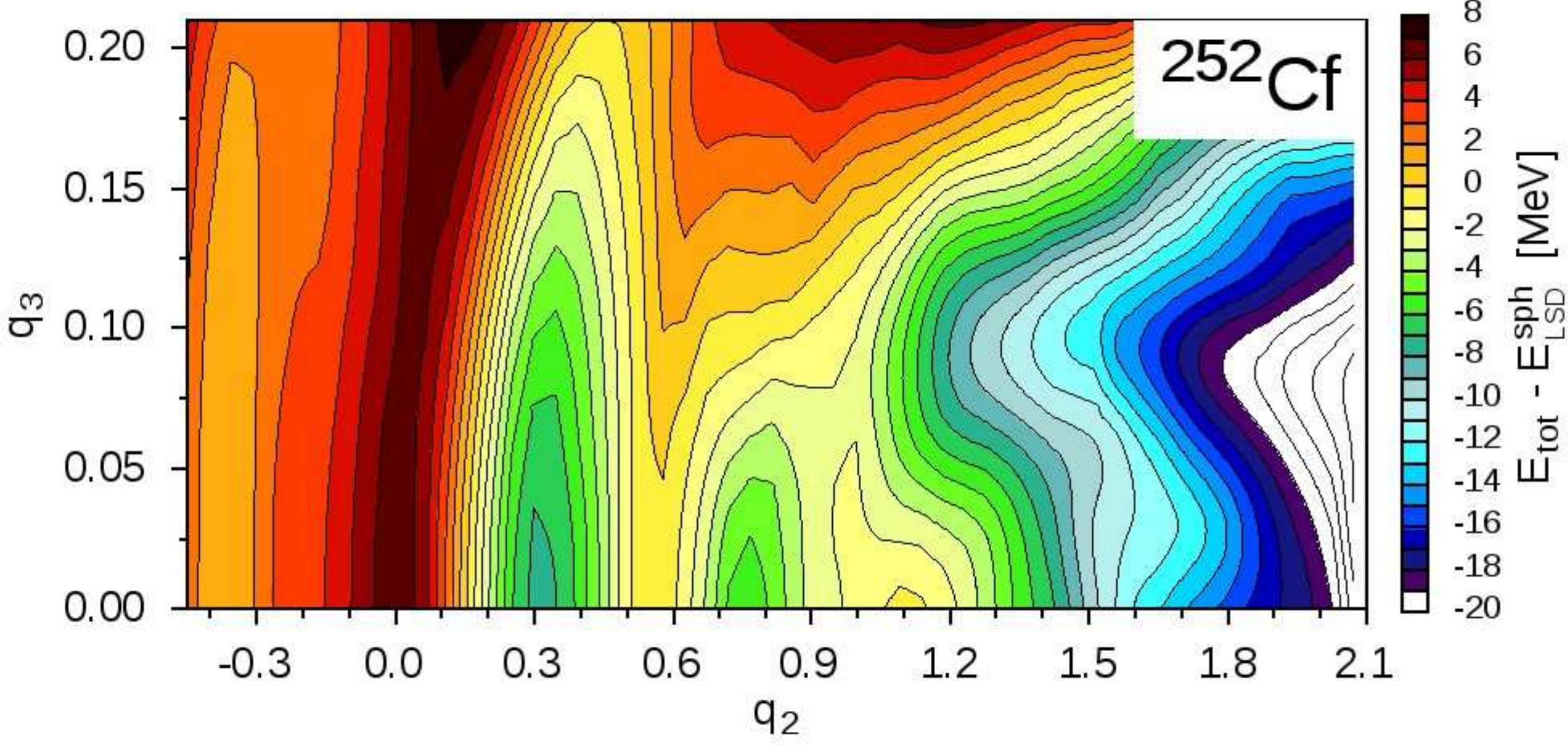}
\includegraphics[width=6.5cm]{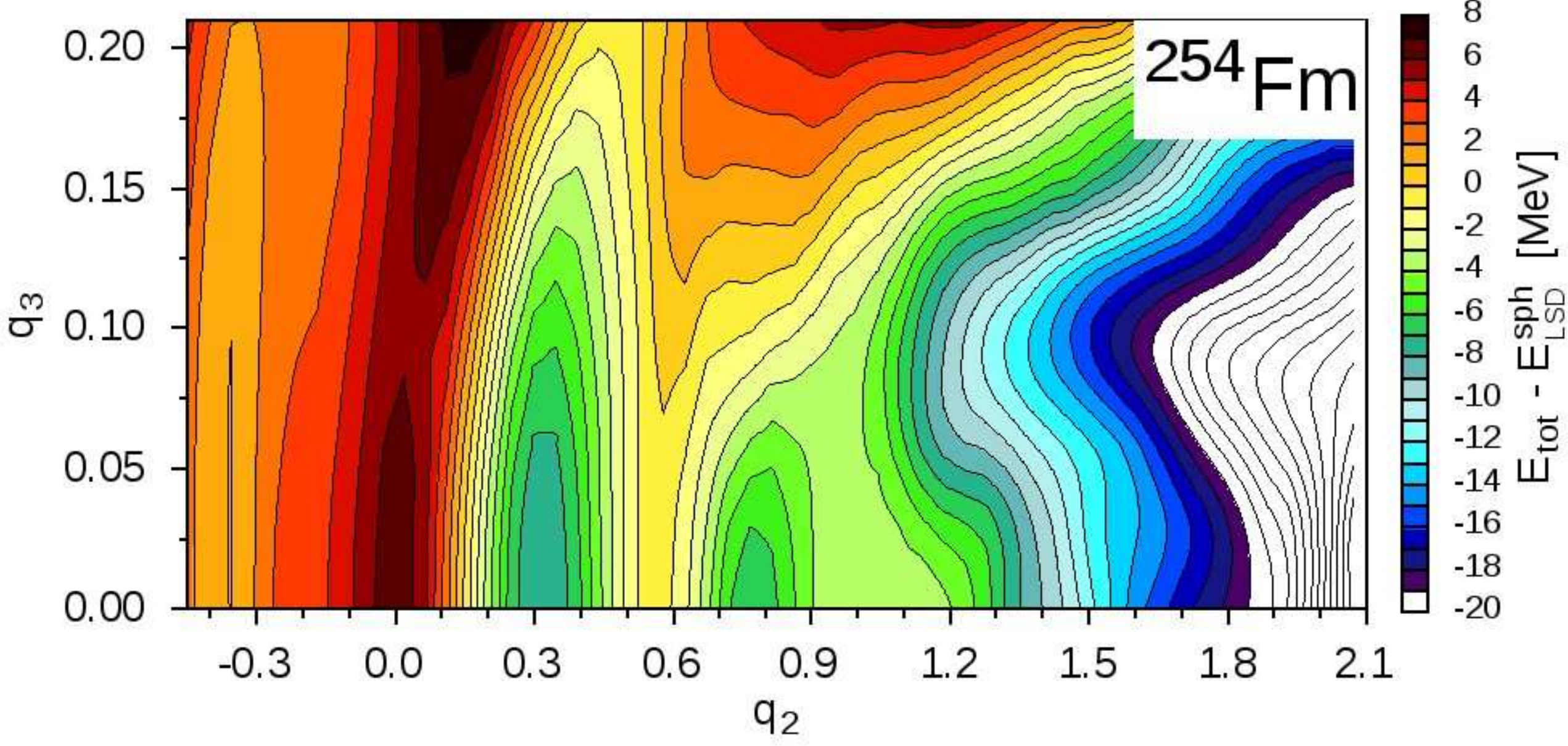}\\[5ex]
\includegraphics[width=6.5cm]{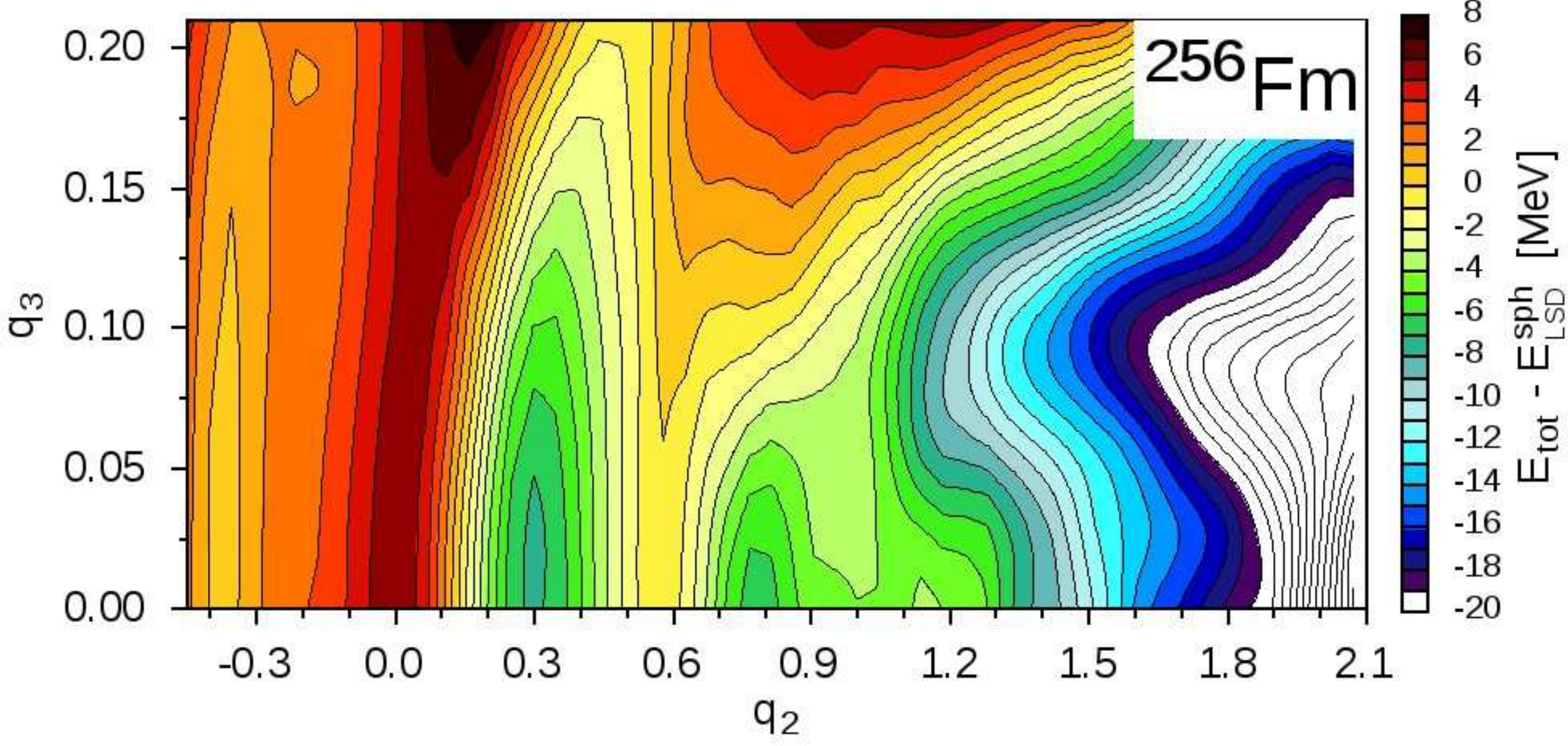}
\includegraphics[width=6.5cm]{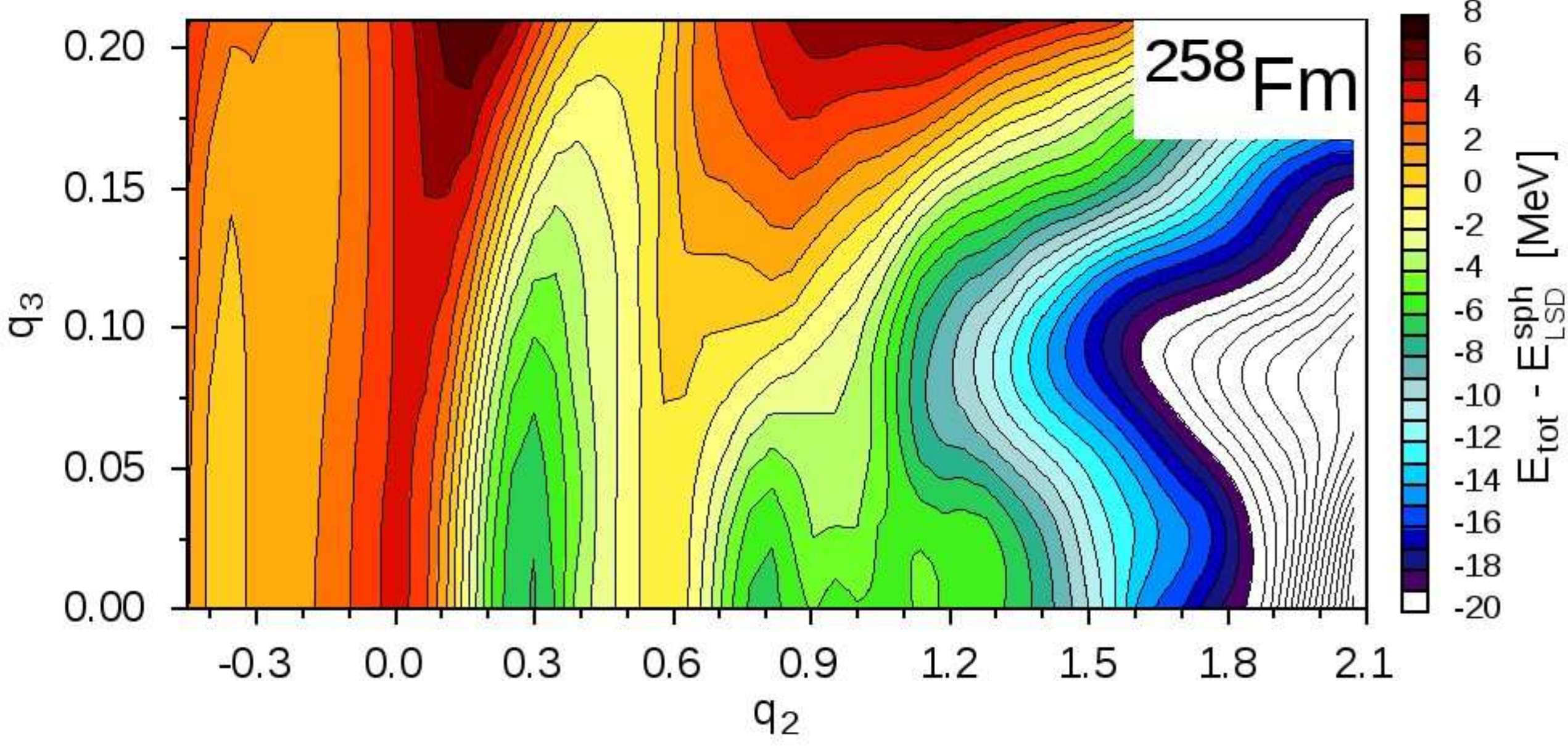}
\caption{(Color online) Deformation energy in the ($q_2, q_3$) plane
minimized with respect to $\eta$ and $q_4$, for $^{252}$Cf and 
$^{254-258}$Fm.}
\label{fig14}\end{figure*}


\subsection{Fission valleys and fission modes}\label{valleys}

Fission-fragment properties (mass, charge, and kinetic-energy distributions) are
determined by the evolution of the fissioning system on the multidimensional
deformation-energy landscape. This path, between the initially compact
configuration up to the scission into two separate fragments, is influenced by
both static and dynamical effects, with the former effects presumably dominating
at low excitation energy. High quality calculations of the potential-energy
landscape are therefore essential for reliable predictions on fragment
properties. Conversely, measured fragment properties, and at first place, their
mass distributions, constitute a unique tool for probing the potential-energy
landscape, and thereby testing the underlying model.
                                                                     \\[ -2.0ex]

It is an experimentally well-established fact that the fission-fragment mass
distribution in low-energy fission of actinides around U is asymmetric (see
\cite{andreyev2018} and Refs.\ therein). On the theoretical side, there is also
general agreement that this asymmetry originates from the influence of shell
effects in the nascent fragments \cite{mosel1971, zhang2016}. A further analysis
of the correlation between mass and kinetic energy reveals the presence of
different fission channels or modes \cite{wilkins1976, brosa1990}. Dominant
fission modes are attributed to the influence of shell effects in the nascent
{\it heavy} fission fragment, one channel near the doubly magic $^{132}$Sn, and
another at a deformed shell closure around neutron number $N \!=\! 88$. The
competition between these two modes and the additional symmetric mode depends
on the fissioning nucleus \cite{dematte1997, schmidt2016}. The sharp transition
from asymmetric to symmetric fission which was experimentally observed between 
$^{256}$Fm and $^{258}$Fm \cite{hulet1986}, with a very narrow mass distribution
and high kinetic energy, was interpreted as the signature of the formation of
two close-to-magic Sn isotopes in a compact scission configuration
\cite{nagame2012}. In other words, with increasing mass of the fissioning 
system, the light-mass peak in actinide fission approaches the heavy one. The
same feature was observed for several isotopes of other elements slightly beyond
Fm \cite{HESS2017, hulet1986, hulet1980, hulet1989, nagame2012}.
                                                                     \\[ -2.0ex]

The experimental findings in the Fm region triggered a very intense theoretical
effort, with both macroscopic-microscopic and self-consistent models in order 
to identify the origin behind this particular fission mechanism (see e.g.
Refs.~\cite{pashkevich1988, moller2001, ichikawa2009, pasca2017, bonneau2006,
warda2002, dubray2008}). In our previous work \cite{SNP17} we analyzed the 4D
potential-energy landscapes computed within the present approach for a wide
range of pre-actinides and actinides. Although only qualitative at this level
\footnote{A more quantitative estimate of the fragment mass distribution would
require dynamical calculations. Static arguments based on the sole
potential-energy landscape remain qualitative, even though, in low-energy
fission, the potential-energy topography allows already for a faithful estimate
of the shape of the distribution.}, the study showed that our model provides a
consistent description of the competition and evolution of asymmetric and
symmetric fission up to Pu. Motivated by this encouraging result, we propose in
the present work to extend the study to heavier elements. It is in particular
interesting to investigate whether the 4D deformation space based on the Fourier
shape parametrization is able to account for the specific modes which lead to
the abrupt transition observed in the Fm region.
                                                                     \\[ -2.0ex]
\begin{figure*}[htb]
\includegraphics[width=6.5cm]{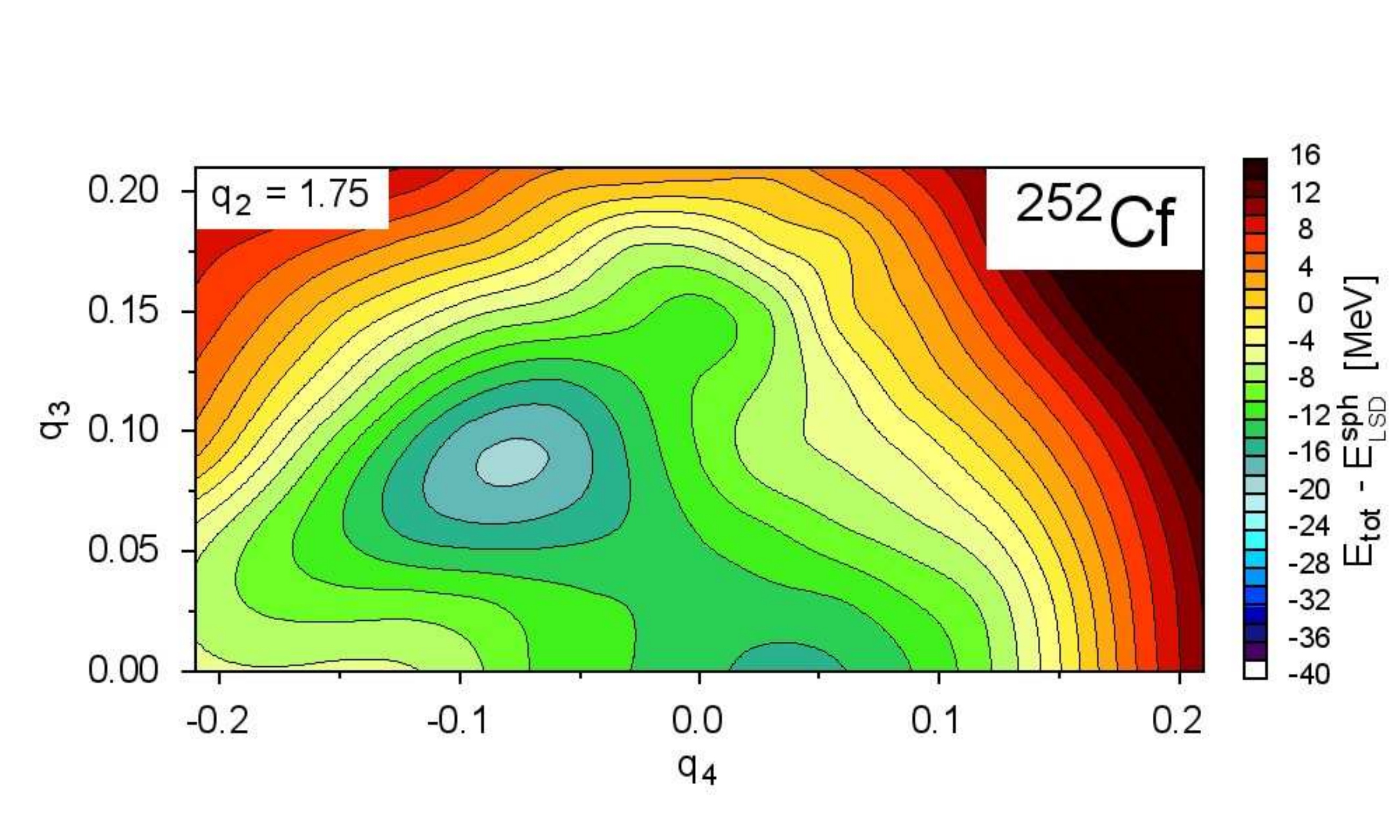}
\includegraphics[width=6.5cm]{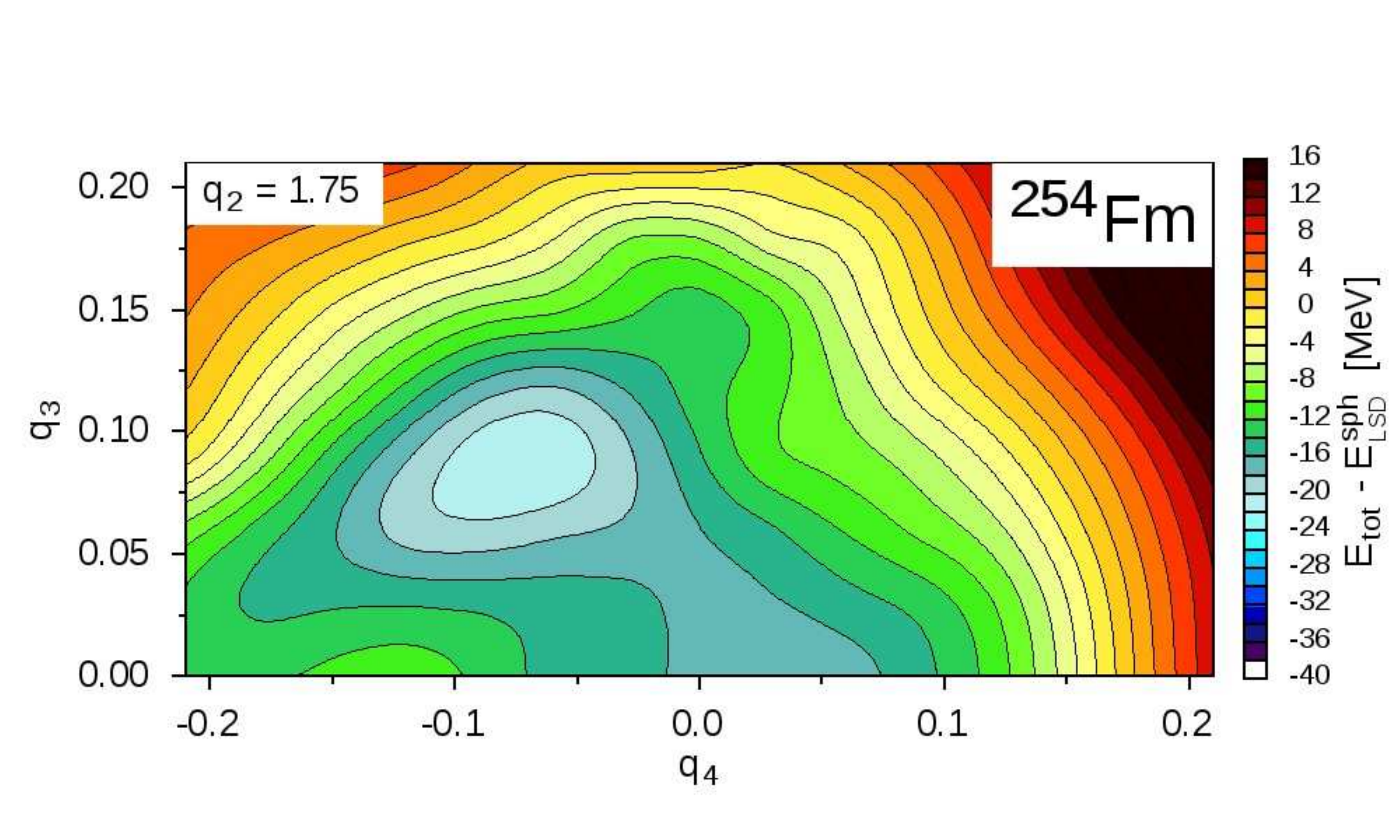}\\
\includegraphics[width=6.5cm]{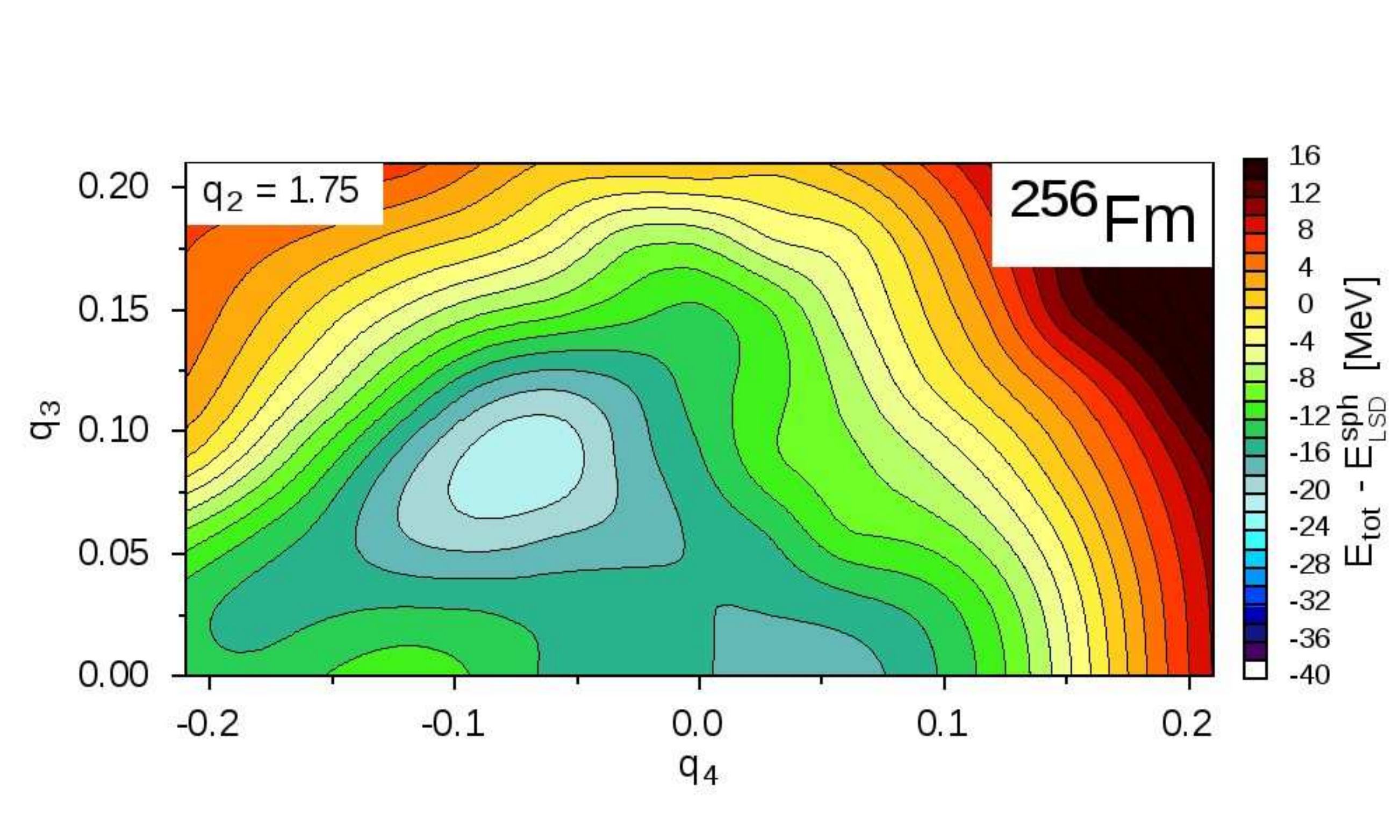}
\includegraphics[width=6.5cm]{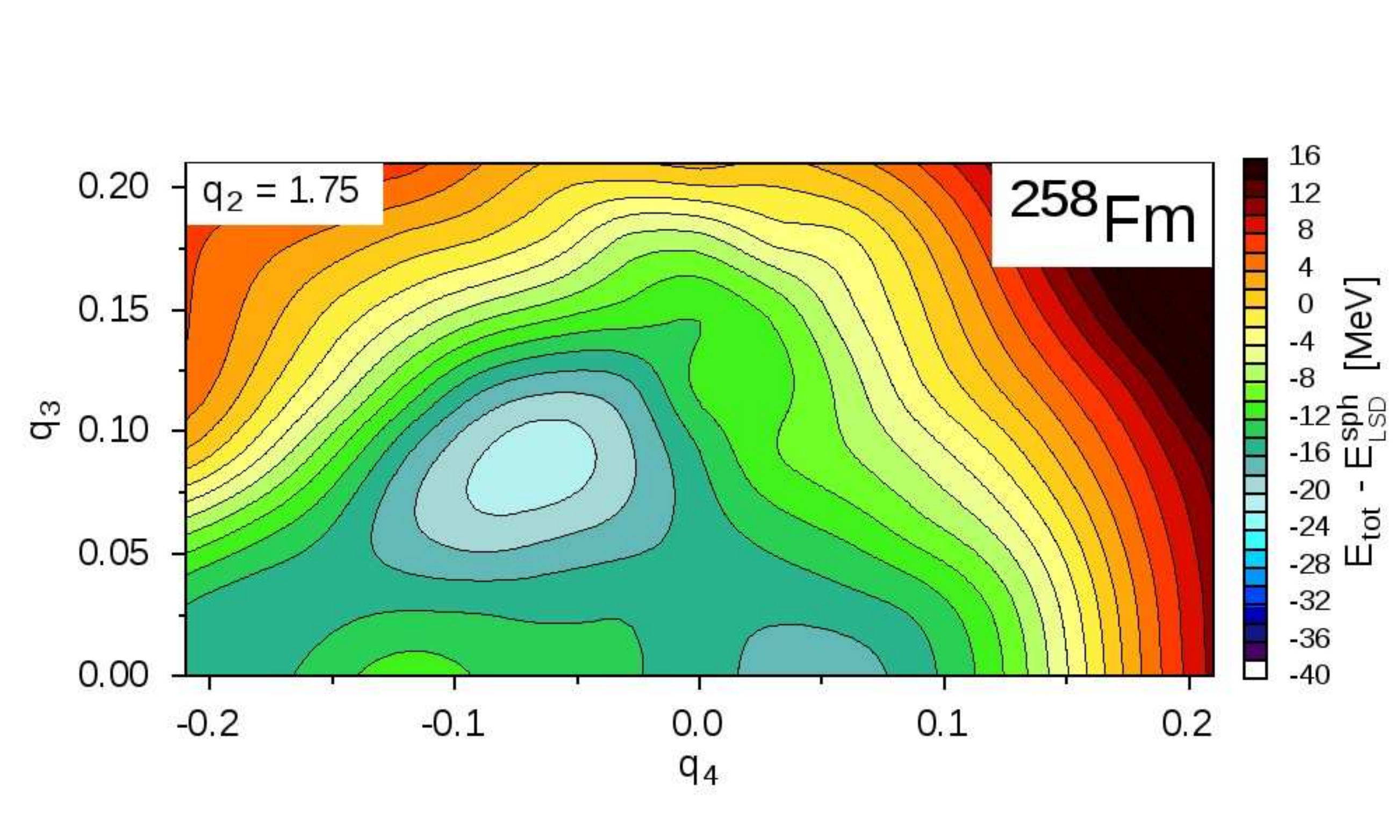}
\caption{(Color online) Deformation energy in the ($q_4, q_3$) plane for $q_2 =
1.75$, for $^{252}$Cf and $^{254-258}$Fm.} 
\label{fig16}\end{figure*}
The method used to identify fission valleys in the 4D landscape was detailed in
Ref.~\cite{SNP17}. In short, we identify as a fission valley a continuous path,
running through the 4D space, with the criterion of slowly varying values along
each of the collective coordinates. As a first step, we search for paths to
fission in the ($q_2$, $q_3$) potential-energy map obtained after minimization
with respect to $\eta$ and $q_4$. Then, along each ``candidate path'' (or
valley), step by step in $q_2$, we look whether it is associated with a
continuous set of minima in the other 2D spaces, like $(\eta,\, q_2)$, $(q_2,\,
q_4)$, $(q_3,\, q_4)$, etc. 
                                                                     \\[ -2.0ex]

The ($q_2,\, q_3$) potential-energy maps for $^{252}$Cf and $^{254-258}$Fm are
displayed in Fig.\ 13. For all these nuclei, one observes a prolate-deformed
ground-state minimum at $q_2 \approx 0.3$ and an isomeric minimum at $q_2
\approx 0.8$. For $^{252}$Cf the topography of the landscape resembles that of
lighter actinides (see e.g.\ the case of $^{228}$Ra in Fig.\ 8 of Ref.\ 
\cite{SNP17}), although some structures at and beyond the outer saddle ($q_2
\approx 1.1$) are somewhat less pronounced. The maximum at the outer saddle
being located around $q_3 = 0$, the ($q_2$, $q_3$) landscape suggests that
$^{252}$Cf will predominantly fission asymmetrically at low excitation energy.
This is confirmed by the ($q_3$, $q_4$) maps analyzed for successive $q_2$ along
the path to scission. In Fig.\ 14 (upper left) we show the map of $^{252}$Cf at
$q_2 = 1.75$. The latter value was selected corresponding to an elongation
beyond the outer-saddle region, where the descent to scission is already
initiated. The choice is somehow arbitrary, but since the fission valley runs
nearly parallel to the $q_2$ axis once the descent is initiated (see Fig.\ 13),
the location in $q_3$ (equivalently, mass asymmetry) of the actual scission 
does not depend strongly on this precise choice of $q_2$. As noted previously,
the left-right asymmetry of the valley is determined rather early, slightly
beyond the outer saddle \cite{mosel1971}. The ($q_4$, $q_3$) map shows two
minima, one at ($q_3 \approx 0.07$, $q_4 \approx -0.075$) and the other at
($q_3 \approx 0.0$, $q_4 \approx 0.03$), corresponding respectively to
asymmetric and symmetric mass splits, with compact and elongated scission 
configurations \cite{SNP17}. The former minimum is much deeper, suggesting the
dominance of mass-asymmetric fission, with a location in $q_3$ corresponding to
a heavy fragment mass around 140, consistent with experiment
\cite{schmitt1966}. Comparing this result in particular with Fig.\ 8 of
Ref.~\cite{SNP17} one concludes that the pattern and location of the fission 
valleys in the 4D landscape for $^{252}$Cf are similar to those obtained for the
lighter actinides.

\begin{figure}[htb]
\includegraphics[width=8.cm,height=7.cm, angle=0]{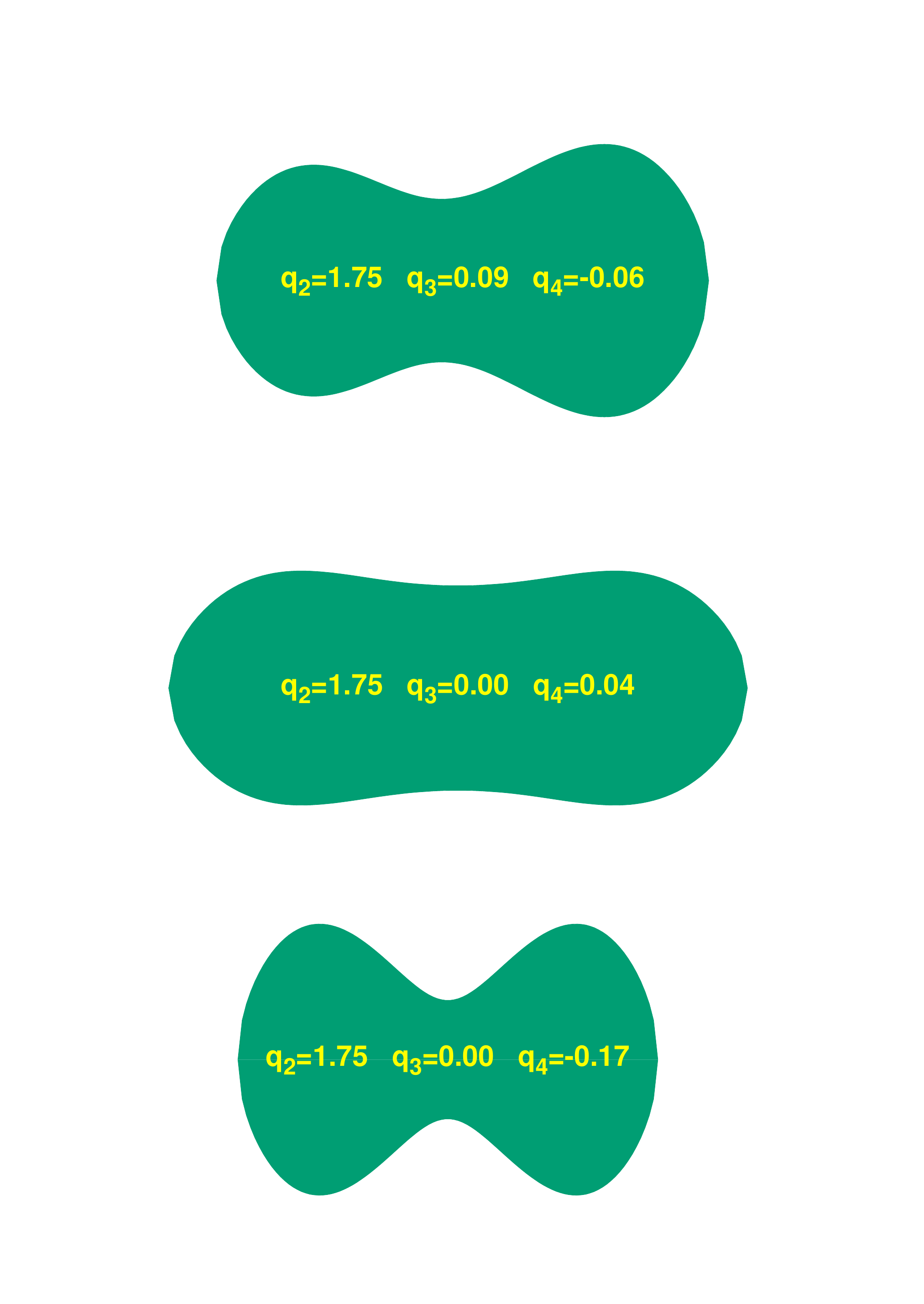}
\caption{(Color online) Nuclear shape for three different sets of collective
variables ($q_2$, $q_3$, $q_4$).} 
\label{fig17}\end{figure}

Moving to the Fm isotopes, a close inspection of Fig.\ 13 suggests a progressive
departure from the above trend with increasing Fm mass. For $^{254}$Fm the
potential-energy landscape in the outer-saddle region ($q_2 \approx 1.1$) is
rather flat in $q_3$ direction with the absence of a maximum centered at $q_3 =
0$ as this was the case for $^{252}$Cf. For heavier Fm isotopes, the potential
energy even decreases towards symmetry. In other words, there is a progressive
migration of the lowest outer-saddle point from ($q_3 \approx 0.07$) to ($q_3
\approx 0.0$) between $^{254}$Fm and $^{258}$Fm, suggesting the emergence of a
favored symmetric splitting. This preference of equal-mass partition increases
further for still heavier Fm isotopes (not shown). In order to get a deeper
insight into the reason for this migration, let us consider the ($q_3$, $q_4$)
maps for $q_2 = 1.75$ in Fig.\ 14. For all three Fm isotopes, one notices the
presence of the same minima as for $^{252}$Cf, identified as the compact
asymmetric and elongated symmetric (so-called super-long) modes. However, with
increasing Fm mass, a {\it third} distinct minimum starts to develop, localized
at $q_3 = 0$ and $q_4$ in the range $-0.2$ to $-0.15$, suggesting a second mode
of symmetric fission. The profiles of the nuclear shapes corresponding to the
three minima identified in Fig.\ 14 are displayed in Fig.\ 15. The two minima,
common to Cf and Fm, are recognized as the compact asymmetric and elongated
symmetric channels, while the third mode emerging for the heaviest Fm is seen to
correspond to a particularly compact symmetric configuration. This result is 
completely in line with the experimental finding \cite{hulet1986}, as well as
with the observations by other models \cite{ichikawa2009, warda2002}, which
confirms the capability of the Fourier shape parametrization to describe the
rich variety of shapes encountered in fission \cite{SNP17}, as well as the
accuracy of the potential-energy calculation.
                                                                     \\[ -2.0ex]

We would like to emphasize at this point that this achievement of our shape
parametrization was {\it a priori} not at all guaranteed, due to the limited
number (4, actually only 3 in the present case, since the nonaxial deformation
$\eta$ plays here a rather minor role) of collective coordinates. This
investigation demonstrates that the Fourier parametrization is indeed able to
generate the appearance of two distinct symmetric fission valleys across the
Fermium region, with properties that are supported by the experimental data. It
shall be noted that the relative strength of the different fission modes cannot
be deduced quantitatively from the present study. To do so, an extended range in
$q_2$ starting from before the outer saddle, and the influence of the dynamics,
have to be taken into account. This is beyond the scope of this work.
                                                                     \\[ -2.0ex]

Encouraged by the above achievement of the model, we propose to extend its
application to still heavier elements. In particular, we are interested in
investigating what happens {\it beyond} the abrupt change evidenced in the heavy
Fm isotopes. Very little experimental information exists only up to Rf
\cite{hulet1989}, due to the difficulty in accessing this region of the nuclear
chart with standard methods.
                                                                     \\[ -2.0ex]

The analysis of the No isotopic chain exhibits a tendency very similar to that
of the Fm one: the prevalence of asymmetric fission slowly changes to favored
symmetric fission between No mass 256 and 262. Similarly to Fm, depending on the
fissioning system, two (for lighter No) or three (for heavier No) distinct
fission paths are predicted. 
                                                                     \\[ -2.0ex]

Along the chain of Rf, the same kind of pattern is again observed. A few ($q_2$,
$q_3$) maps were shown in Fig.\ 8. The ($q_4$, $q_3$) maps projected at $q_2 =
1.75$ are displayed in Fig.\ 16 for two isotopes. Like for Fm and No, a second
compact symmetric fission path appears for the heavier isotopes. Notice however
that the outer barrier decreases below 2 MeV, when it does not nearly vanish for
the heaviest Rf isotopes displayed (see $^{264}$Rf in Fig.\ 8). The barrier
disappears completely when going to still higher $Z$ numbers. In addition, as
seen for some Hs isotopes in Fig.\ 8, the potential-energy landscape along the
descent to scission can be rather flat in the $q_3$ direction. In spite of that
softness, well-localized and well-defined valleys persist. This is demonstrated
for $^{264}$Rf and $^{272}$Hs in the ($q_4$, $q_3$) maps of Fig.\ 16, where
three distinct minima are again predicted. One concludes from this analysis
that in order to localize these different fission valleys, it is absolutely
essential to analyze the multidimensional energy landscape, as we have done
here, and that a pure consideration of say the ($q_2$, $q_3$) map would have
failed to evidence these different fission valleys.
                                                                     \\[ -2.0ex]

Interestingly, it is also observed that the minimum corresponding to the
elongated symmetric scission configuration tends to disappear with increasing Hs
mass, and this, in favor of the compact symmetric configuration. This result
suggests a new change in the picture of fission above the Fm chain. Below Fm,
there is a coexistence between an asymmetric compact and a symmetric elongated
mass split. In the Fm region, an additional contribution from a symmetric
compact configuration emerges. For heavier elements, the pattern transforms into
the coexistence of an asymmetric and a symmetric partition, where both of these 
correspond to a compact shape. To our knowledge the question of the evolution of
fission modes beyond the Fm transition is addressed here for the first time. The
new feature predicted by our model arises from the complex structure of the
multidimensional potential-energy landscape. It would be very interesting to
investigate our predictions both theoretically (with other models), and
experimentally as a next probe of the richness of available deformation spaces
and the precision of potential-energy calculations. Robust and accurate
predictions are indeed of prime importance for a further extension of fission
calculation in the SHE region. 
\begin{figure}[!h]
\begin{center}
\hspace{-0.5cm}
\includegraphics[width=9.00cm, height=6.80cm,angle=0]{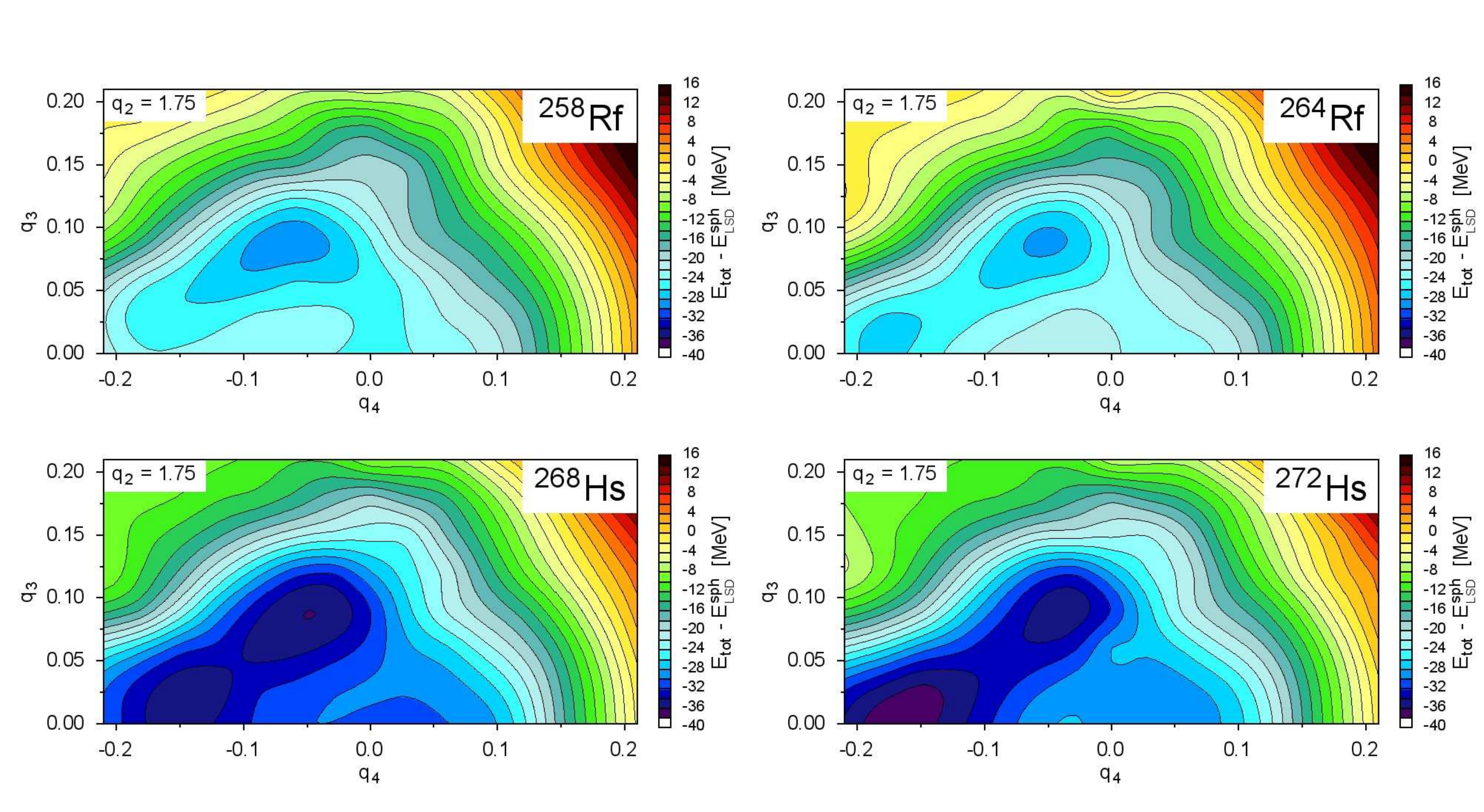}
\caption{(Color online) Deformation energy as in Fig.\ 14 at $q_2 =
1.75$, for $^{258,264}$Rf and $^{268,272}$Hs.} 
\end{center}
\label{fig18}\end{figure}

\section{Conclusions}\label{summary}

The recently developed four-dimensional Fourier parametrization of nuclear
shapes, combined with the extensively tested and successful
macroscopic-microscopic approach of the potential energy based on the
Lublin-Strasbourg Drop and microscopic shell and pairing corrections, is
employed to predict the properties of very heavy and super-heavy nuclei. A
careful analysis of the 4D potential-energy landscapes of 324 even-even isotopes
with $92 \le Z \le 126$ and isospins $40 \le N \!-\!Z \le 74$ allows to study
the evolution with proton and neutron number of the equilibrium ground-state
configuration, the possible presence of isomers, the properties of ground-state 
$\alpha$-radioactivity ($Q_\alpha$-value and half-life) and of spontaneous
fission (barrier heights and fission half-lives). The results of our
calculations have proven to reproduce the experimental data for all these
observables to a good accuracy, whenever such data were available. The enhanced
stability of SHE with $N = 162$, and to a lesser extend for $N = 152$, is, in
particular, clearly evidenced in our model results.
                                                                     \\[ -2.0ex]

These calculations anticipate that nearly all investigated isotopes in the
discussed region are characterized in the ground-state by axial and left-right symmetry. In general, nuclei with $A \leq 280$ are prolate, those with $280
\le A \le 300$ are spherical or transitional, while for $Z \geq$ 122 
and $A \!\geq\! 300$ oblate configuration tends to dominate. For the majority 
of nuclei with $Z \!\leq\! 104$ prolate-deformed shape isomers are predicted.
                                                                     \\[ -2.0ex]

The results by our model suggest that the next shell closure will appear in the
vicinity of $Z \!\!=\!\! 116$ and $N \!=\! 174$ with a rather wide island of
relative stabilization from $Z \approx 114$ to $Z \approx 118$ with $48 \le N
\!-\!Z \le 66$.
                                                                     \\[ -2.0ex]

The present theoretical framework is finally employed to investigate the
evolution across the Fm region of the properties of possibly multiple fission
paths. The abrupt change in the fragment-mass and kinetic energy distributions
observed experimentally between $^{256}$Fm and $^{258}$Fm, and attributed to the
appearance of ``double shell-stabilized'' symmetric splits, is consistent with
the calculated emergence in the 4D deformation space of a fission valley leading
to a compact symmetric scission configuration. The model anticipates a change in
the fission mode picture beyond the Fm transition. Shell-stabilized asymmetric
and symmetric channels are predicted to dominate, whereas the macroscopic-driven
symmetric (super-long) partition tends to disappear. The evolution of fission
modes {\it beyond} Fm is addressed here for the first time. Our conjecture of
shell-dominated channels for fission in the vicinity of Hs would be interesting
to investigate further, from the theoretical and the experimental point of
view.
                                                                     \\[ -2.0ex]
 
Altogether, the here presented achievement of the Fourier shape parametrization 
in a 4-dimensional deformation space, combined with reliable potential energy
calculations offers an attractive basis for dynamical calculations related
both with collective rotational and vibrational excitations as well as with
fission. \\[2.0ex]


{\bf Acknowledgements}\\

This work has been partly supported by the Polish-French COPIN-IN2P3
collaboration agreement under project number 08-131 and by the Polish
National Science Center, grant No. 2016/21/B/ST2/01227.

\end{document}